\documentclass[12pt]{article}
\pdfoutput=1

\usepackage{textcomp}
\usepackage{color}

\usepackage{putex}
\usepackage{autobreak}
\usepackage{graphicx}
\graphicspath{{plots/}}
\usepackage{tabularx}
\usepackage{caption}
\usepackage{amsmath}
\usepackage{array}
\usepackage[font=small]{subcaption}
\usepackage{epstopdf}
\usepackage{enumerate}
\usepackage{cite}
\usepackage{youngtab}
\usepackage{tensor}
\usepackage{slashed}
\usepackage[aligntableaux=center]{ytableau}
\usepackage[utf8]{inputenc}
\usepackage{rotating}
\usepackage{multirow}
\usepackage[
      colorlinks=true,
      linkcolor=blue,
      urlcolor=blue,
      filecolor=black,
      citecolor=red,
      ]{hyperref}
\usepackage{bm} 
\usepackage{booktabs}

\newcommand{\grp}[1]{\mathrm{#1}}

\newcommand{\grU}{\grp{U}}

\newcommand {\be} {\begin {equation}}
\newcommand {\ee} {\end {equation}}

\newcommand {\bes} {\begin {equation*}}
\newcommand {\ees} {\end {equation*}}

\newcommand{\es}[2] {\begin{equation} \label{#1} \begin{split} #2 \end{split} \end{equation}}

\newcommand{\R}{\mathbb{R}}

\newcommand{\cS}{{\mathcal S}}

\newcommand{\beq}{\begin{equation}}
\newcommand{\eeq}{\end{equation}}

\def\p{\partial}

\def\ie{\begin{equation}\begin{aligned}}
\def\fe{\end{aligned}\end{equation}}

\numberwithin{equation}{section}

\def\<{\langle}
\def\>{\rangle}

\graphicspath{{figures/}}

\begin{document}

\preprint{}

\institution{Imp}{Blackett Laboratory, Imperial College, Prince Consort Road, London, SW7 2AZ, U.K.}
\institution{PU}{Joseph Henry Laboratories, Princeton University, Princeton, NJ 08544, USA}
\institution{PCTS}{Princeton Center for Theoretical Science, Princeton University, Princeton, NJ 08544, USA}

\title{
Higher-derivative corrections in M-theory from precision numerical bootstrap
}

\authors{Shai M.~Chester,\worksat{\Imp} Ross Dempsey,\worksat{\PU} and Silviu S. Pufu\worksat{\PU,\PCTS}  }

\abstract{We study higher-derivative corrections to the graviton scattering amplitude in M-theory, via the stress tensor correlator of 3d $\mathcal{N} =8$ $\grU(N)_k\times \grU(N)_{-k}$ ABJM theory (dual to graviton scattering in M-theory on $\text{AdS}_4\times S^7/\mathbb{Z}_k$). We use the conformal bootstrap combined with an integral constraint derived from supersymmetric localization in order to constrain semishort OPE coefficients appearing in the stress tensor correlator. We obtain islands that are significantly more precise than those in previous studies that did not use the integral constraint. Using these islands, we can estimate the powers and coefficients in a large central charge expansion. This allows us to accurately read off the N$^3$LO contribution, from the protected $D^6 R^4$ correction, and also estimate the N$^4$LO contribution, from the unprotected $D^8 R^4$ correction.
}
\date{December 2024}

\maketitle

\tableofcontents

\section{Introduction and Summary}
\label{intro}

M-theory, the eleven-dimensional unification of superstring theories, is hard to study because it is inherently strongly-coupled. We can, however, expand the graviton scattering amplitude in M-theory at small momentum in units of the inverse Planck length, $\ell_{11}^{-1}$. As we will describe in more detail later (see Section~\ref{sec:large_ct}), the leading contribution to this amplitude comes from the Einstein-Hilbert term, given by an integral of the Ricci scalar $R$ along with its supersymmetric completion, in the low-energy effective action. Subleading contributions come from higher-derivative corrections to the Einstein-Hilbert term as well as from loop diagrams.

The first two higher-derivative interactions in the Lagrangian density are of the schematic form $R^4$ (meaning a specific contraction of four Riemann tensors) and $D^6 R^4$ (likewise, a contraction of four Riemann tensors and six derivatives), along with their respective supersymmetric completions. They give rise to protected contributions to the graviton scattering amplitude, which have been derived in various papers \cite{Green:1997as, Russo:1997mk, Green:2005ba,Alday:2020tgi} by using the duality to Type IIA string theory when the eleventh direction is small. The next term in the Lagrangian density, which is of the schematic form $D^8 R^4$, is the leading unprotected term. Its contributions to a given observable cannot be fixed using the duality to Type IIA, and thus require a new non-perturbative approach.

Such an approach is made possible by the holographic relationship between M-theory on AdS$_4\times S^7/\mathbb{Z}_k$ and a 3d $\mathcal{N} = 8$ superconformal field theory (SCFT) introduced by Aharony, Bergman, Jafferis, and Maldacena, called $\grU(N)_k\times \grU(N)_{-k}$ ABJM theory \cite{Aharony:2008ug}, where $k = 1,2$. In terms of the stress tensor two-point function coefficient $c_T$ of this CFT, which scales as $c_T\sim N^{3/2}$ at large $N$ \cite{Drukker:2010nc}, the holographic dictonary states that
\es{cPlanck}{
  \frac{L^6}{\ell_{11}^6}=\left(\frac{3\pi c_T k}{2^{11}}\right)^{\frac23}+O(c_T^0) \,,
}
where $L$ is the AdS$_4$ radius. In particular, the large-$N$ or large-$c_T$ limit of the CFT corresponds to the small curvature limit (in Planck units) in the bulk. We will also study the $\grU(N)_2\times \grU(N+1)_{-2}$ theory of Aharony-Bergman-Jafferis (ABJ) \cite{Aharony:2008gk}, which also has $\mathcal{N} = 8$ supersymmetry.

In \cite{Chester:2018aca}, it was suggested to study the low-energy effective action of M-theory by using the holographic relationship between graviton scattering in the bulk and the stress tensor four-point function in the CFT. One can use analytic bootstrap methods \cite{Rastelli:2017udc,Zhou:2017zaw} to express this stress tensor correlator at any given order in the derivative expansion in terms of a finite number of coefficients. These coefficients can in principle be written in terms of the CFT data. Afterwards, using the methods of \cite{Penedones:2010ue}, one can take the flat-space limit of the AdS holographic correlator and obtain the flat-space graviton scattering amplitude of M-theory.

In \cite{Chester:2018aca}, the contribution of an $R^4$ vertex was fixed using the analytic bootstrap in terms of one unknown coefficient, which was then determined using supersymmetric localization \cite{Kapustin:2009kz}. Likewise, in \cite{Binder:2018yvd}, the contribution of a $D^4 R^4$ vertex was fixed in terms of two unknown coefficients. As we will explain in Section~\ref{sec:constraints}, at a fixed value of $c_T$ we have two independent constraints from supersymmetric localization, and so again this case could be handled analytically (and it was shown that $D^4 R^4$ is absent in M-theory, consistent with expectations from string dualities).\footnote{See \cite{Binder:2019mpb,Binder:2021cif} for similar results in the type IIA limit and higher spin gravity limits.}

Starting with the $D^6 R^4$ term, the two known constraints from supersymmetric localization are no longer sufficient. In the contribution from the $D^6 R^4$ term there are three unknown coefficients; the contribution from the $D^8 R^4$ term (the leading unprotected term) has four unknown coefficients; and so on. Thus, we need new inputs. In this paper, we will describe progress towards using the numerical conformal bootstrap \cite{Rattazzi:2008pe} to combine the localization constraints with further non-perturbative constraints from unitarity,\footnote{Unitarity does not impose nontrivial constraints in the large-$N$ expansion of the analytic bootstrap, because OPE coefficients squared are trivially positive in the generalized free field theory that describes the strict large-$N$ limit.} thus extending our ability to study M-theory beyond the supergravity limit using holography.

The numerical conformal bootstrap for the stress tensor correlator in 3d $\mathcal{N}=8$ SCFTs such as ABJ(M) theory was initiated in \cite{Chester:2014fya,Chester:2014mea}, where it was used to obtain bounds on CFT data. The CFT data accessible from the stress tensor correlator comprises the OPE coefficients of the short and semishort multiplets that can be exchanged, and the scaling dimensions and OPE coefficients of the long multiplets that can be exchanged. The most stringent bounds obtained were islands giving the feasible region in the space of OPE coefficients of semishort multiplets. 

In \cite{Agmon:2017xes,Agmon:2019imm}, these bounds were significantly improved by using one constraint from localization to compute the OPE coefficients of short multiplets analytically \cite{Kapustin:2009kz,Nosaka:2024gle}, and inputting their values at a given $N$ and $k$ into the bootstrap. After this step, the bounds on semishort OPE coefficients were tight enough to match the known large-$c_T$ expansion at order $c_T^{-1}$, which corresponds to the $R$ vertex in the bulk, and at order $c_T^{-2}$, which corresponds to a one-loop term with two $R$ vertices in the bulk. However, the bounds were not yet tight enough to be meaningfully compared with the next term in the large-$c_T$ expansion, which corresponds to the $D^6 R^4$ vertex in the bulk---let alone the subsequent term, which corresponds to the $D^8 R^4$ vertex.

\begin{figure}[t]
	\centering
	\includegraphics[width=.6\linewidth]{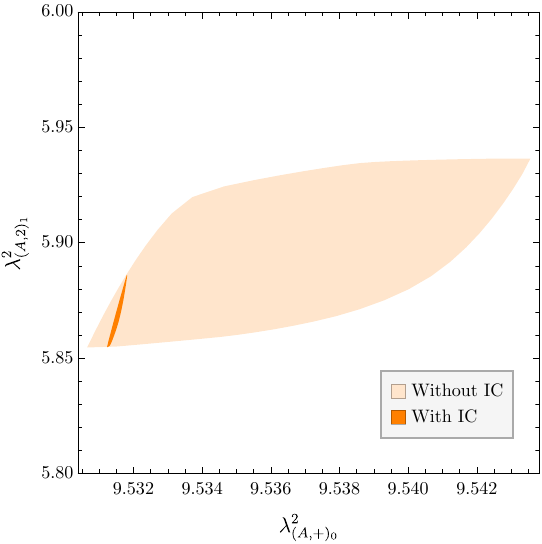}
	\caption{An example of the numerical bootstrap bounds in the space of the squared OPE coefficients $\lambda^2_{(A,+)_0}$ and $\lambda^2_{(A,2)_1}$ for the interacting sector of the $\grU(2)_1\times \grU(2)_{-1}$ ABJM theory. The island without the additional localization constraint, first computed in \cite{Agmon:2017xes,Agmon:2019imm}, is already quite small. By including a localization constraint, in this paper we reduce the size of this island by a factor of about 200.}
	\label{fig:island_n2_compare}
\end{figure}

In this paper, we improve further upon the numerical bootstrap bounds on semishort OPE coefficients by imposing the second constraint from supersymmetric localization. This constraint relates a certain integral of the stress-tensor correlator to a derivative of the mass-deformed sphere free energy $F(m_\pm)$, namely $\partial_{m_+}^2\partial_{m_-}^2F(m_\pm)\vert_{m_\pm=0}$. It was recently shown in \cite{Chester:2021aun,Chester:2023ehi} how to combine integral constraints of this form with the numerical bootstrap in the context of 4d $\mathcal{N}=4$ super-Yang-Mills (SYM). We generalize this approach to the 3d ABJM bootstrap, and find that the resulting bounds are significantly more constrained than those of \cite{Agmon:2017xes,Agmon:2019imm}. As an example, in Figure~\ref{fig:island_n2_compare} we show how the island in the space of certain semishort OPE coefficients squared is reduced by a factor of about 200 relative to the islands reported in \cite{Agmon:2017xes,Agmon:2019imm}. 

We compare these extremely precise islands at large $N$ to analytic predictions \cite{Zhou:2017zaw,Chester:2018lbz,Chester:2018aca,Binder:2018yvd,Alday:2021ymb,Alday:2022rly}. If we only input the known large-$c_T$ expansions up to order $c_T^{-2}$, we find that we can extract the power and coefficient of the next term, and that it matches the known $D^6 R^4$ contribution at order $c_T^{-7/3}$ \cite{Alday:2021ymb,Alday:2022rly}. Following the same method, we input the known expansion of semishort OPE coefficients to $\mathcal{O}(c_T^{-7/3})$, and extract the correct power and a tentative estimate of the coefficient of the subsequent contribution, coming from the unprotected $D^8 R^4$ vertex.

The rest of this paper is organized as follows. In Section~\ref{4point} we review the superblock decomposition of the stress tensor four-point correlator, the constraints on the correlator from crossing symmetry and supersymmetric localization, and the structure and known terms of the large-$c_T$ expansion of the CFT data. In Section~\ref{sec:bootstrap}, we discuss how to use the integral constraint from localization in tandem with the numerical bootstrap, and we use this approach to compute bounds on CFT data for a wide range of $N$ and $k$. We then compare these bounds to the analytic large-$c_T$ results. Finally, in Section~\ref{disc} we end with a discussion of our results and of future directions. Various technical details are discussed in the Appendices.

\section{Stress tensor four-point function}
\label{4point}

The main object of study in this paper is the stress tensor multiplet four-point function. We begin by discussing general constraints coming from the $\mathfrak{osp}(8|4)$ superconformal group, including the expansion in superblocks. We then give the exact constraints on the correlator that relate either some protected OPE coefficients or a certain integrals of the correlator to derivatives of the mass deformed sphere free energy, which can be efficiently computed using supersymmetric localization. Finally, we discuss the large-$c_T$ expansion of the semishort OPE coefficients $\lambda^2_{(A,+)_0}$ and $\lambda^2_{(A,2)_1}$, which we will compare to our numerical bounds obtained for finite $N$ in the next section.

The material in Sections~\ref{setup} and \ref{sec:constraints} follows the setup given in \cite{Chester:2014fya,Agmon:2017xes,Agmon:2019imm}. More details can be found in these references. We review key results here in order to fix notation and provide a concise summary of facts needed for this paper. Likewise, in Section~\ref{sec:large_ct}, we review results first obtained in \cite{Zhou:2017zaw,Chester:2018lbz,Chester:2018aca,Binder:2018yvd,Alday:2021ymb,Alday:2022rly} in order to provide the explicit large-$c_T$ expansions that we will need in Section \ref{sec:results}.

\subsection{Setup}
\label{setup}

Graviton scattering in M-theory on AdS$_4\times S^7/\mathbb{Z}_k$ is captured by the four-point function of the stress tensor in 3d ABJ(M) theory. The stress tensor belongs to a supermultiplet with a superconformal primary denoted $S$, which is a dimension-one scalar in the $\bm{35}_c = [0020]$ irrep of the $\mathfrak{so}(8)$ R-symmetry algebra. Thus, to study graviton scattering, we can study the four-point function $\langle SSSS\rangle$.

Since the $\bm{35}_c$ is the rank-two traceless symmetric tensor irrep, it is convenient to define
\begin{equation}
    S(x,Y) = S_{IJ}(x)Y^I Y^J\,,
\end{equation}
where $Y^I$ with $I=1,\ldots,8$ is an auxiliary polarization vector with $Y\cdot Y = 0$. Superconformal invariance constrains the four-point function to have a block expansion of the form \cite{Dolan:2004mu}
\es{SBDecomp}{
   \langle S( x_1,Y_1)S( x_2,Y_2)S( x_3,Y_3)S( x_4,Y_4) \rangle=\frac{(Y_1\cdot Y_2)^2(Y_3\cdot Y_4)^2}{| x_{12}|^{2}| x_{34}|^{2}}\sum_{\mathcal{M}\in\{S\times S\}}\lambda^2_\mathcal{M}\mathfrak{G}_\mathcal{M}(U,V;\sigma,\tau)\,.
}
Here $\lambda_{\mathcal{M}}^2$ are squared OPE coefficients for the fusion of two stress tensor supermultiplets into another supermultiplet $\mathcal{M}$, and $\{S\times S\}$ is shorthand for the set of such supermultiplets. These supermultiplets are grouped into an $A$-series and a $B$-series, which have different types of shortening conditions; the supermultiplet denoted $(A,0)$ is a long multiplet with no shortening conditions. More details on the representations of $\mathfrak{osp}(8|4)$ can be found in \cite{Dolan:2008vc}, and the supermultiplets in $\{S\times S\}$ along with the $\mathfrak{so}(3,2)\oplus\mathfrak{so}(8)\subset \mathfrak{osp}(8|4)$ quantum numbers of their superconformal primaries are given in Table~\ref{tab:opemult}.

 The cross-ratios $U, V, \sigma$, and $\tau$ are given by
\es{uvsigmatauDefs}{
  U \equiv \frac{{x}_{12}^2 {x}_{34}^2}{{x}_{13}^2 {x}_{24}^2} \,, \qquad
   V \equiv \frac{{x}_{14}^2 {x}_{23}^2}{{x}_{13}^2 {x}_{24}^2}  \,, \qquad
   \sigma\equiv\frac{(Y_1\cdot Y_3)(Y_2\cdot Y_4)}{(Y_1\cdot Y_2)(Y_3\cdot Y_4)}\,,\qquad \tau\equiv\frac{(Y_1\cdot Y_4)(Y_2\cdot Y_3)}{(Y_1\cdot Y_2)(Y_3\cdot Y_4)} \,,
 }
with $x_{ij} \equiv x_i - x_j$. With this choice of $\sigma$ and $\tau$, the superconformal Ward identities imply that the correlator is a quadratic polynomial, which we write as
\es{4point3d}{
\langle S(\vec x_1,Y_1)S(\vec x_2,Y_2)S(\vec x_3,Y_3)S(\vec x_4,Y_4)\rangle =\frac{(Y_1\cdot Y_2)^2(Y_3\cdot Y_4)^2}{| x_{12}|^{2}| x_{34}|^{2}}\Big[\cS_1(U,V)+\sigma^2&\cS_2(U,V)\\
+\tau^2\cS_3(U,V)+\sigma\tau \cS_4(U,V)+\tau\cS_5(U,V)+\sigma&\cS_6(U,V)\Big]\,.\\
}

\begingroup
\renewcommand{\arraystretch}{1.3}
\begin{table}
\centering
\begin{tabular}{>{\centering\arraybackslash}m{2cm}>{\centering\arraybackslash}m{3cm}>{\raggedleft\arraybackslash}m{3cm}>{\centering\arraybackslash}m{2cm}>{\centering\arraybackslash}m{3cm}}
\toprule
Type & $(\Delta,\ell)$ & $\mathfrak{so}(8)$ irrep & spin $\ell$ & OPE coefficient \\ 
\midrule
\text{Id} &  $(0,0)$  & $\mathbf{1} = [0000]$ & 0 & $\lambda_{\text{Id}}$  \\ 
$(B,+)$ &  $(1,0)$  & $\mathbf{35}_c = [0020]$  & 0 & $\lambda_{\text{Stress}}$  \\ 
$(B,+)$ &  $(2,0)$  & $\mathbf{294}_c = [0040]$ & 0 & $\lambda_{(B,+)}$ \\ 
$(B,2)$ &  $(2,0)$  & $\mathbf{300} = [0200]$ & 0 & $\lambda_{(B,2)}$ \\ 
$(A,+)$ &  $(\ell+2,\ell)$  & $\mathbf{35}_c = [0020]$ & even & $\lambda_{(A,+)_\ell}$ \\ 
$(A,2)$ &  $(\ell+2,\ell)$  & $\mathbf{28} = [0100]$ & odd & $\lambda_{(A,2)_\ell}$ \\ 
$(A,0)$ &  $\Delta\ge \ell+1$  & $\mathbf{1} = [0000]$ & even & $\lambda_{(A,0)_{\Delta,\ell}}$  \\ 
\bottomrule
\end{tabular}
\caption{The possible superconformal multiplets in the $S \times S$ OPE.  The $\mathfrak{so}(3, 2) \oplus \mathfrak{so}(8)$ quantum numbers are those of the superconformal primary in each multiplet.}
\label{tab:opemult}
\end{table} 
\endgroup

When we write the OPE coefficients, we use the notation given in Table~\ref{tab:opemult}. In the conventions of \cite{Chester:2014fya}, if we normalize $S$ such that the OPE coefficient of the identity operator is $\lambda_{\text{Id}} = 1$, then
\es{cTRel}{
  \lambda_{\text{Stress}}^2 \equiv  \frac{256}{c_T} \,,
 }
where $c_T$ is the coefficient appearing in the two-point function of the canonically normalized stress tensor:
 \es{CanStress}{
  \langle T_{\mu\nu}(\vec{x}) T_{\rho \sigma}(0) \rangle = \frac{c_T}{64} \left(P_{\mu\rho} P_{\nu \sigma} + P_{\nu \rho} P_{\mu \sigma} - P_{\mu\nu} P_{\rho\sigma} \right) \frac{1}{16 \pi^2 \vec{x}^2} \,, \qquad P_{\mu\nu} \equiv \eta_{\mu\nu} \nabla^2 - \partial_\mu \partial_\nu \,.
 }
Here, $c_T$ is normalized such that it equals $1$ for a theory containing just a (non-supersymmetric) free massless real scalar, or just a free massless Majorana fermion. Thus, $c_T = 16$ for the free ${\cal N} = 8$ theory of eight massless real scalars and eight massless Majorana fermions.

\subsection{Bootstrap constraints}\label{sec:constraints}

In this paper, we will focus on bounding the OPE coefficients of the semishort multiplets $(A,+)_\ell$ (with $\ell$ even) and $(A,2)_\ell$ (with $\ell$ odd). To do this, we will use constraints from the crossing symmetry of the four-point function \eqref{SBDecomp} in combination with additional constraints from supersymmetric localization.

The four-point function \eqref{SBDecomp} is non-trivially invariant under swapping the first operator with the third. The resulting crossing equations were derived in \cite{Chester:2014fya}; here we review the results and give the independent crossing equations we will use in Section~\ref{sec:bootstrap}.

The superblocks $\mathfrak{G}_\mathcal{M}(U,V;\sigma,\tau)$ can be expanded into channels of the bosonic subgroup $\mathfrak{so}(3,2)\oplus \mathfrak{so}(8)$. For the $\mathfrak{so}(8)$ R-symmetry factor, we use the decomposition
\begin{equation}
	\bm{35}_c \otimes \bm{35}_c = \bm{1} \oplus \bm{28} \oplus \bm{35}_c \oplus \bm{294}_c \oplus \bm{300} \oplus \bm{567}_c\,.
\end{equation}
The contributions of these irreps multiply the following polynomials in $\sigma$ and $\tau$ \cite{Nirschl:2004pa,Chester:2014fya}:
\begin{equation}\label{eq:y_polys}
    \begin{aligned}
        Y_{\bm{1}} &= 1, & Y_{\bm{28}} &= \sigma - \tau, & Y_{\bm{300}} &= \sigma^2 + \tau^2 - 2\sigma\tau - \frac{1}{3}(\sigma+\tau) + \frac{1}{21},\\
        Y_{\bm{35}_c} &= \sigma + \tau - \frac{1}{4}, & Y_{\bm{567}_c} &= \sigma^2 - \tau^2 - \frac{2}{5}(\sigma-\tau), & Y_{\bm{294}_c} &= \sigma^2 + \tau^2 + 4\sigma\tau - \frac{2}{3}(\sigma+\tau) + \frac{1}{15}\,.
    \end{aligned}
\end{equation}
For the $\mathfrak{so}(3,2)$ factor, the contribution of a conformal multiplet with a primary of scaling dimension $\Delta$ and spin $\ell$ multiplies the 3d conformal block $G_{\Delta,\ell}(U,V)$, which we normalize as in \cite{Chester:2014fya}. The expansion of the superblock then takes the form
\begin{equation}\label{eq:superblock}
    \mathfrak{G}_{\mathcal{M}}(U,V;\sigma,\tau) = \sum_{\bm{r}\in \bm{35}_c\otimes\bm{35}_c} Y_{\bm{r}}(\sigma,\tau)\sum_{\Delta,\ell} \mathfrak{a}^{\mathcal{M}}_{\bm{r},\Delta,\ell} G_{\Delta,\ell}(U,V).
\end{equation}
The explicit coefficients $\mathfrak{a}^\mathcal{M}_{\bm{r},\Delta,\ell}(\Delta,\ell)$ are given in Appendix~C of \cite{Chester:2014fya}, as well as in a Mathematica notebook attached to \cite{Agmon:2019imm}. 

As explained in \cite{Chester:2014fya}, the crossing symmetry of the four-point function can be reduced to the two equations\footnote{Na\"ively there are six crossing equations, but due to superconformal Ward identities, four of these are redundant.}
\begin{equation}\label{eq:crossing_explicit}
\begin{split}
    V^{(1)}(U,V) &= \sum_{\mathcal{M}\in\{S\times S\}} \lambda^2_\mathcal{M} V^{(1)}_\mathcal{M} = 0\,,\\
    V^{(2)}(U,V) &= \sum_{\mathcal{M}\in\{S\times S\}} \lambda^2_\mathcal{M}  V^{(2)}_\mathcal{M} = 0\,,
\end{split}
\end{equation}
where the functions $V^{(1)}$ and $V^{(2)}$ are given by
\begin{equation}
\begin{split}
    V^{(1)}_\mathcal{M} &= \sum_{\Delta,\ell} \left(a^{\mathcal{M}}_{\bm{28},\Delta,\ell} + a^{\mathcal{M}}_{\bm{35}_c,\Delta,\ell} + \frac{5}{3}a^{\mathcal{M}}_{\bm{300},\Delta,\ell} - \frac{2}{5}a^{\mathcal{M}}_{\bm{567}_c,\Delta,\ell} - \frac{14}{3}a^{\mathcal{M}}_{\bm{294}_c,\Delta,\ell}\right)F^{\Delta,\ell}_+(U,V)\,,\\
    V^{(2)}_\mathcal{M} &= \sum_{\Delta,\ell} \left(a^{\mathcal{M}}_{\bm{1},\Delta,\ell} - \frac{1}{4} a^{\mathcal{M}}_{\bm{35}_c,\Delta,\ell} - \frac{20}{21}a^{\mathcal{M}}_{\bm{300},\Delta,\ell}  + a^{\mathcal{M}}_{\bm{567}_c,\Delta,\ell} - \frac{14}{15}a^{\mathcal{M}}_{\bm{294}_c,\Delta,\ell}\right)F^{\Delta,\ell}_+(U,V)\,,
\end{split}
\end{equation}
and the function $F^{\Delta,\ell}_+(U,V)$ is defined as
\begin{equation}
    F_+^{\Delta,\ell}(U,V) \equiv V^2 G_{\Delta,\ell}(U,V) + U^2 G_{\Delta,\ell}(V,U)\,.
\end{equation}

In addition to these crossing equations, there are additional constraints coming from supersymmetric localization. As explained in \cite{Chester:2014fya,Closset:2012vg,Closset:2012ru}, we can fix $c_T = 256/\lambda^2_\text{Stress}$ by using localization to relate it to the free energy of the mass-deformed theory on $S^3$, denoted $F(m_+, m_-)$. The exact relationship is
\begin{equation}\label{eq:m2der}
  c_T = \frac{64}{\pi^2}\frac{\partial^2 F}{\partial m_\pm^2} \bigg|_{m_\pm=0}  \,.
\end{equation}
Additionally, as explained in \cite{Agmon:2017xes,Dedushenko:2016jxl}, using four derivatives with respect to mass we can compute the following linear combination of short OPE coefficients:
\begin{equation}\label{eq:m4der}
-\frac{2^{13}}{\pi^4 c_T^2} \left.\frac{\partial^4 F}{\partial m_\pm^4}\right|_{m_\pm = 0} = -4+\frac12{ \lambda}^2_{(B,+)} +\frac18{ \lambda}^2_{(B,2)} \,.
\end{equation}
Furthermore, in \cite{Chester:2014mea}, the following linear relationship is derived:
\begin{equation}\label{eq:crossConstraints}
    4\lambda^2_\text{Stress} - 5\lambda^2_{(B,+)} + \lambda^2_{(B,2)} + 16 = 0\,.
\end{equation}
Thus, provided we can compute the free energy derivatives, we have enough information to fix these three OPE coefficients. Recently in \cite{Nosaka:2024gle}, the mass-deformed sphere free energy for $\grU(N)_k\times \grU(N+M)_{-k}$ ABJ(M) theory was computed exactly for any $N$, $M$, $k$, and $m_\pm$ in terms of a recursion relation. In Appendix~\ref{app:localization}, we review this recursion relation and give example values of $\lambda^2_\text{Stress}$, $\lambda^2_{(B,+)}$, and $\lambda^2_{(B,2)}$. The sphere free energy can also be computed to all orders in $1/N$ using the Fermi gas method \cite{Marino:2011eh,Nosaka:2015iiw}, which we also describe in Appendix~\ref{app:localization}.

The third and final known constraint from localization, derived in \cite{Binder:2018yvd,Binder:2019mpb}, is the main addition to the numerical approach of this paper relative to \cite{Agmon:2017xes,Agmon:2019imm}. In terms of the expansion \eqref{4point3d}, it takes the form 
\es{con3d2}{
-\frac{64}{\pi^2c_T^2}\left.\frac{\partial^4 F}{\partial m_+^2 \partial m_-^2}\right|_{m_\pm = 0} &= I[\cS^i_\text{conn.}]\,,
}
where
\begin{equation}\label{eq:integral_operator}
I[\cS^i]\equiv\frac{1}{96}\int dR\,d\theta\,\sin\theta\left[\frac{\cS^1(U,V)}{U}+\frac{\cS^2(U,V)}{U^2}+\frac{V\cS^3(U,V)}{U^2}\right]\bigg|_{\substack{U = 1 + R^2 - 2 R \cos \theta \\
    V = R^2}}\,,
\end{equation}
By $\cS^i_\text{conn.}$ we mean the connected part of the correlator: 
\es{3dcon}{
\vec\cS_\text{conn.}=\vec\cS-\begin{pmatrix}1&U&\frac{U}{V}&0&0&0\end{pmatrix}\,.
}
Note that the integrand in \eqref{eq:integral_operator} is crossing symmetric.

\subsection{Large-$c_T$ expansion}\label{sec:large_ct}

As described in Section~\ref{sec:constraints}, the OPE coefficients $\lambda^2_\text{Stress}$, $\lambda^2_{(B,+)}$, and $\lambda^2_{(B,2)}$ of the short multiplets can all be computed exactly. In Section~\ref{sec:results}, we will describe the use of the numerical conformal bootstrap to constrain the remaining CFT data. As we will show in Section~\ref{sec:results}, our constraints for the OPE coefficients of the semishort multiplets, $\lambda^2_{(A,+)_\ell}$ and $\lambda^2_{(A,2)_\ell}$, are by far the most precise. We will thus aim to study higher-derivative corrections in M-theory via the large-$c_T$ expansions of these coefficients.

These corrections are all diffeomorphism-invariant terms appearing in the effective 11-dimensional supergravity action. This action is given by $S = S_\text{EH} + S_\text{HD}$, where
\begin{equation}
    S_\text{EH} \propto \ell_{11}^{-9} \int d^{11}x\,R + \text{supersymmetric completion}
\end{equation}
and $S_\text{HD}$ contains higher-derivative corrections. The first several terms contributing to four-point graviton scattering that are allowed by the supersymmetric Ward identities take the schematic form
\begin{equation}
    \ell_{11}^{-3} R^4, \quad\ell_{11} D^4 R^4,\quad \ell_{11}^{3} D^6 R^4,\quad \ell_{11}^{5} D^8R^4, \ldots\,.
\end{equation}
Here $D^m R^n$ denotes a specific contraction of $m$ derivatives with $n$ Riemann tensors, and we have multiplied by the power of the Planck length required by dimensional analysis. As discussed in \cite{Chester:2018aca}, the $R^4$ contact term does not contribute to the semishort OPE coefficients in ABJM theory. In addition, the $D^4 R^4$ term is absent from the M-theory effective action \cite{Green:1999pu,Green:2005ba,Binder:2018yvd}. Thus, for the purposes of this paper, we will be focused on the $D^6 R^4$ and the $D^8 R^4$ terms.

The scaling in $c_T$ and $k$ of the various contributions to the four-graviton scattering amplitude can be determined as follows. For the tree-level terms, the holographic dictionary \eqref{cPlanck} tells us that $\ell_{11} \propto (c_T k)^{-1/9}$. This means that the coefficient of $R$ in the effective four-dimensional theory, which is proportional to $\ell_{11}^{-9} \Vol(S^7/\mathbb{Z}_k)$, is in turn proportional to $c_T$ and independent of $k$. We can then read off the scalings of the higher-derivative coefficients simply by dividing by the appropriate power of $\ell_{11}$. This means, for instance, that the coefficients of the $D^6 R^4$ and $D^8 R^4$ terms in the effective 4d theory are proportional to $k^{-4/3} c_T^{-7/3}$ and $k^{-14/9} c_T^{-23/9}$, respectively. From these terms, one can compute contact Witten diagrams that have this same scaling.

The scaling of one-loop diagrams is more complicated. The first loop contribution has two $R$ vertices, and enters at order $c_T^{-2}$; the next has an $R$ vertex and an $R^4$ vertex, and enters at order $c_T^{-8/3}$. These loop diagrams involve a sum over Kaluza-Klein modes on $S^7/\mathbb{Z}_k$, and so they do not have a simple power-law scaling in $k$.

In ABJ(M) theory, we are working in a normalization convention where the two-point function of the operator $S$ in the correlator \eqref{SBDecomp} is independent of $k$ and $c_T$, and so we can read off the structure of the large-$c_T$ expansion from the scaling of Witten diagrams contributing to the graviton scattering amplitude. In summary, the structure of the large-$c_T$ expansion for the semishort OPE coefficients in ABJ(M) theory is as follows:
\begin{equation}
\begin{split}
    \lambda^2 &= \lambda^2_0 + \lambda^2_R c_T^{-1} + \lambda^2_{R|R} f_{R|R}(k) c_T^{-2} + \lambda^2_{D^6 R^4} k^{-4/3} c_T^{-7/3} \\
    &\qquad + \lambda^2_{D^8 R^4} k^{-14/9} c_T^{-23/9} + \lambda^2_{R|R^4} f_{R|R^4}(k) c_T^{-8/3} + \cdots\,.
\end{split}
\end{equation}
This expansion has been computed to $O(c_T^{-7/3})$ in \cite{Zhou:2017zaw,Chester:2018lbz,Chester:2018aca,Binder:2018yvd,Alday:2021ymb,Alday:2022rly}. In particular, for the $D^6 R^4$ term, there are enough constraints from supersymmetry to analytically fix its contribution. However, the $D^8 R^4$ term is unprotected, and so we do not know the $\mathcal{O}(c_T^{-23/9})$ contribution to the semishort OPE coefficients.

For the two semishort OPE coefficients we will consider in Section~\ref{sec:bootstrap}, the large-$c_T$ expansion is \cite{Zhou:2017zaw,Chester:2018lbz,Chester:2018aca,Binder:2018yvd,Alday:2021ymb,Alday:2022rly}
\es{strong}{
&k=1:\quad \lambda^2_{(A,+)_0}=
\frac{64}{9}
-\frac{1
   0240 \left(2 \pi ^2-21\right)}{27 \pi ^2
  c_T}
  +\frac{513.492}{{c_T}^2}
-\frac{917504\cdot ({\frac{2}{3}})^{\frac13}   }{9 \pi ^{10/3}
   {c_T}^{7/3}}+O(c_T^{-23/9})\,,\\
& \qquad\qquad\;\,  \lambda^2_{(A,2)_1}=
\frac{1024}{105}
-\frac{16384 \left(48 \pi ^2-455\right)}{315 \pi ^2
  c_T}
  +\frac{5221.36}{{c_T}^2}
-\frac{4194304\cdot 18^{\frac13}   }{5 \pi ^{10/3}
   {c_T}^{7/3}}+O(c_T^{-23/9})\,,\\
&k=2: \quad \lambda^2_{(A,+)_0}=\frac{64}{9}
-\frac{1
   0240 \left(2 \pi ^2-21\right)}{27 \pi ^2
  c_T}
  +\frac{285.320}{{c_T}^2}
-\frac{458752\cdot ({\frac{1}{3}})^{\frac13}   }{9 \pi ^{10/3}
   {c_T}^{7/3}}+O(c_T^{-23/9})\,,\\
   &  \qquad\qquad \,\;\lambda^2_{(A,2)_1}=
\frac{1024}{105}
-\frac{16384 \left(48 \pi ^2-455\right)}{315 \pi ^2
  c_T}
  +\frac{2239.90}{{c_T}^2}
-\frac{2097152\cdot 3^{\frac23}   }{5 \pi ^{10/3}
   {c_T}^{7/3}}+O(c_T^{-23/9})\,.\\
}

This expansion appears to converge quickly, even for relatively low values of $N\propto c_T^{2/3}$. We will see this in Section~\ref{sec:bootstrap}, but as some initial evidence for it, we can take the large-$c_T$ expansion of the short OPE coefficients $\lambda^2_{(B,2)}$ and $\lambda^2_{(B,+)}$, and compare with the exact values obtained using localization. For example, the large-$c_T$ expansion for $\lambda^2_{(B,2)}$ is \cite{Zhou:2017zaw,Chester:2018lbz,Chester:2018aca,Binder:2018yvd,Alday:2021ymb,Alday:2022rly}
\es{strong2}{
&k=1:\quad \lambda^2_{(B,2)}=
\frac{32}{3}
-\frac{1024 \left(4 \pi
   ^2-15\right) }{9 \pi ^2c_T}
   +\frac{40960 \cdot 2^{1/3}
   }{3^{2/3}\pi ^{8/3}c_T^{5/3}}
   -\frac{1024 \left(400-353 \pi ^2+45
   \pi ^4\right) }{9 \pi ^4c_T^2}\\
&\qquad\qquad\qquad\qquad-\frac{655360
   \cdot 2^{2/3} }{3^{4/3} \pi
   ^{10/3}c_T^{7/3}}
+\frac{40960 \cdot 2^{1/3} \left(368+45
   \pi ^2\right) }{9 \cdot 3^{2/3} \pi ^{14/3}c_T^{8/3}}+O(c_T^{-3})\,,\\
   &k=2:\quad \lambda^2_{(B,2)}=
\frac{32}{3}
-\frac{1024 \left(4 \pi
   ^2-15\right) }{9 \pi ^2c_T}
   +\frac{40960 \cdot 2^{-1/3}
   }{3^{2/3}\pi ^{8/3}c_T^{5/3}}
   +\frac{16384 \left(-25+2 \pi ^2\right) }{9 \pi ^4c_T^2}\\
&\qquad\qquad\qquad\qquad-\frac{327680
   \cdot({\frac{2}{3}})^{\frac13} }{3 \pi
   ^{10/3}c_T^{7/3}}
+\frac{7536640 \left(\frac{2}{3}\right)^{2/3}  }{9 \pi ^{14/3}c_T^{8/3}}+O(c_T^{-3})\,,\\
}
and that of $\lambda^2_{(B,+)}$ can be readily obtained from \eqref{eq:crossConstraints}. 

\begin{figure}[t]
	\centering
	\begin{subfigure}[t]{0.48\linewidth}%
		\centering
		\includegraphics[width=\linewidth]{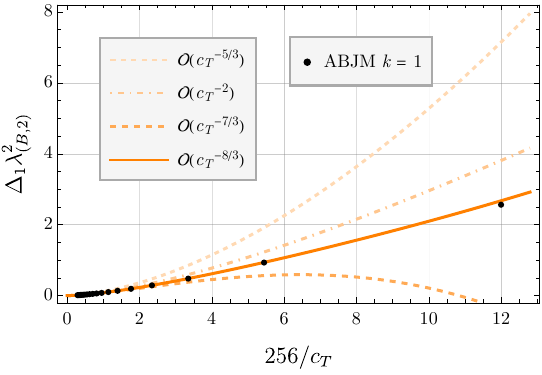}
		\caption{The values for the $k = 1$ ABJM theory.}
	\end{subfigure}%
	\hspace{.02\linewidth}%
	\begin{subfigure}[t]{0.48\linewidth}%
		\centering
		\includegraphics[width=\linewidth]{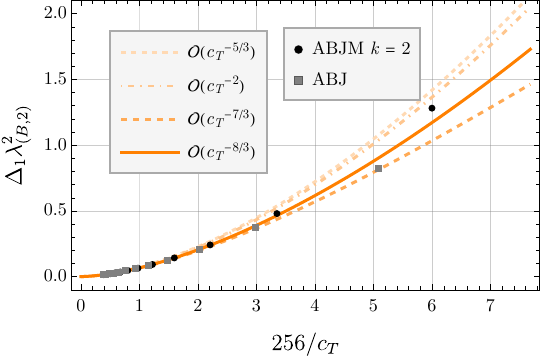}
		\caption{The values for the $k = 2$ ABJM and ABJ theories.}
	\end{subfigure}
	\caption{The squared OPE coefficient $\lambda^2_{(B,2)}$ (with the $\mathcal{O}(1)$ and $\mathcal{O}(1/c_T)$ contributions subtracted off) computed using the strong-coupling expansion \eqref{strong2} at various truncations, compared with the exact values obtained using the recursion relations derived in \cite{Nosaka:2024gle}.}
	\label{fig:b2_cT_expansion}
\end{figure}

In Figure~\ref{fig:b2_cT_expansion} we compare this expansion to the exact values. To be able to see the effects of the higher-order terms in \eqref{strong2}, we plot the residual after subtracting off the $\mathcal{O}\left(c_T^{-1}\right)$ contribution:
\begin{equation}
    \Delta_1 \lambda^2_{(B,2)} \equiv \lambda^2_{(B,2)} - \left(\frac{32}{3}
-\frac{1024 \left(4 \pi
   ^2-15\right) }{9 \pi ^2c_T}\right).
\end{equation}
We see that adding the subsequent terms steadily improves the match between the expansion and the exact values.

\section{Numerical bootstrap}\label{sec:bootstrap}

We now use the numerical bootstrap along with the localization constraints described in Section \ref{sec:constraints} to bound the CFT data of $\mathcal{N}=8$ ABJ(M) at finite $N$ and $k$. We will first describe how to combine the constraints of crossing with the integral constraint, following a similar discussion for 4d $\mathcal{N}=4$ SYM in \cite{Chester:2021aun,Chester:2023ehi}, except now applied to the 3d $\mathcal{N}=8$ bootstrap. We will then present our bootstrap bounds and compare them to the large-$c_T$ expansion given in Section~\ref{sec:large_ct}.

\subsection{Bootstrap algorithm}\label{sec:bootstrap_setup}

Our objective is to obtain upper and lower bounds on the values of semishort OPE coefficients, in particular $\lambda^2_{(A,+)_0}$ and $\lambda^2_{(A,2)_1}$, for the $\grU(N)_k\times \grU(N+M)_{-k}$ ABJ(M) theory for a given $N\ge 2$, $k\in\{1,2\}$, and $0\leq M\leq k-1$. In its primal formulation, this entails solving an optimization problem of the following form:
\begin{equation}\label{eq:primal_problem}
\begin{aligned}
    & \text{min/max} && \lambda^2_{\mathcal{M}^*} \\
    & \text{s.t.} &&
    \begin{aligned}[t]
        & V^{(1)}(U, V) = 0, \\
        & V^{(2)}(U, V) = 0, \\
        & I[\mathcal{S}_\text{conn}] = -\frac{64}{\pi^2 c_T^2} \left.\frac{\partial^4 F}{\partial m_+^2 \partial m_-^2}\right|_{m_\pm = 0},
    \end{aligned} \\
    & \text{with} && \lambda^2_{\mathcal{M}} \geq 0 \quad \text{for all } \mathcal{M}.
\end{aligned}
\end{equation}
Note that the crossing equations $V^{(1)} = V^{(2)} = 0$ (given explicitly in \eqref{eq:crossing_explicit}) include the known OPE coefficients $\lambda^2_\text{Stress}$, $\lambda^2_{(B,+)}$, and $\lambda^2_{(B,2)}$. We input these values. All other OPE coefficients are treated as variables in the optimization problem.

The crossing equations entail an infinite number of linear constraints, parametrized by the values of $U$ and $V$. For the numerical bootstrap, we truncate this set by expanding the crossed blocks $F_+^{\Delta,\ell}(U,V)$ appearing in $V^{(1)}$ and $V^{(2)}$ around the crossing-symmetric point $z = \bar{z} = \frac{1}{2}$, where $U = z\bar{z}$ and $V=(1-z)(1-\bar{z})$:
\begin{equation}\label{eq:F_expansion}
    F^{\Delta,\ell}_{+}(U,V)=\sum_{\substack{p+q\text{ even}\\p\leq q}}\frac{2}{p!q!}\left(z-\frac12\right)^p\left(\bar z-\frac12\right)^q\partial^p_z\partial_{\bar z}^q F^{\Delta,\ell}_{+}(U,V)\vert_{z=\bar z=\frac12}
\end{equation}
We then have constraints
\begin{equation}
    \sum_{\mathcal{M}} \lambda^2_\mathcal{M} V^{(r)}_{\mathcal{M};p,q} = 0\,,
\end{equation}
where by $V^{(r)}_{\mathcal{M};p,q}$ we mean $V^{(r)}_{\mathcal{M}}$ with $F^{\Delta,\ell}_+(U,V)$ replaced by $\partial^p_z\partial_{\bar z}^q F^{\Delta,\ell}_{+}(U,V)\vert_{z=\bar z=\frac12}$, for any choice of $r=1,2$ and $(p,q)$ with $p\leq q$ and $p+q$ even. In \cite{Agmon:2017xes}, it is shown that a linearly independent set of these constraints is given by
\begin{equation}\label{eq:independent_crossing}
    \left\lbrace V^{(1)}_{p> 0,0} = 0, V^{(2)}_{p,q} = 0\right\rbrace\,,
\end{equation}
where by $V^{(1)}_{p> 0,0}$, we mean we only consider derivatives of the first crossing equation with respect to $z$, and we do not include $V^{(1)}_{0,0}$. To get a finite set of constraints, we further restrict to
\begin{equation}
    p+q\leq\Lambda
\end{equation}
for some cutoff $\Lambda$.

In order to impose the integral constraint \eqref{con3d2} in the numerical bootstrap, we need to write $I[\mathcal{S}_\text{conn}]$ explicitly as a linear combination of OPE coefficients (analogously to the expression of the crossing equations in \eqref{eq:crossing_explicit}). That is, we need to determine how the integral operator \eqref{eq:integral_operator} acts upon the superblocks in the expansion \eqref{SBDecomp}. 

The integral in \eqref{con3d2} is over $\R^2$ (written in polar coordinates $(R, \theta)$), while the block expansion of a four-point function has a finite radius of convergence. As discussed in the 4d case \cite{Chester:2021aun,Chester:2023ehi}, we divide $\R^2$ into three regions that are permuted under the $S_3$ crossing symmetry of the four-point function. In terms of the radial coordinates $r,\eta$ defined in \cite{Hogervorst:2013sma} as
\es{retatozzb}{
U=\frac{16 r^2}{\left(r^2+2 \eta  r+1\right)^2}\,,\qquad V=\frac{\left(r^2-2 \eta  r+1\right)^2}{\left(r^2+2 \eta 
   r+1\right)^2}\,,
}
we define the region $D_1$ in which the $s$-channel block expansion converges by
\es{fundDomain}{
D_1:\qquad r\leq -\sqrt{\eta ^2 + 4 | \eta | ++3}+| \eta | +2\,,\qquad |\eta|\leq1\,.
}
We then change variables in \eqref{con3d2} to $r,\eta$, restrict the integration range to $D_1$, and multiply by three, so that the integral acts as
\es{Ireta}{
I[\mathcal{S}_\text{conn}]=\frac{1}{128}\int_{D_1}dr\,d\eta\,\frac{1-r^2}{r^2}\Bigg[\frac{1+r^2+2r\eta}{1+r^2-2r\eta}\mathcal{S}^1(r,\eta)&+\frac{(1+r^2+2r\eta)^5}{256r^2(1+r^2-2r\eta)}\mathcal{S}^2(r,\eta)\\
&+\frac{1+r^2-2r\eta}{1+r^2+2r\eta}\mathcal{S}^3(r,\eta)\Bigg]\,.
}
The functions $\mathcal{S}^i(r,\eta)$ are the ones appearing in \eqref{4point3d}. Using \eqref{eq:y_polys}, these can be related to the basis used for the superblocks in \eqref{eq:superblock}. The superblocks are written as linear combinations of conformal blocks $G_{\Delta,\ell}(U,V)$, and these can be expanded in small $r$ as
\es{3dblockNorm}{
G_{\Delta,\ell}(r,\eta)= r^\Delta\sum_{p=0}^\infty \sum_{s=\max(\ell-p,0)}^{p+\ell} B_{p,s}(\Delta,\ell)r^p P_s(\eta)\,,
}
where $P_s(\eta)$ is a Legendre polynomial and the coefficients $B_{n,s}(\Delta,\ell)$ can be computed to a given order $n$ using the Zamolodchikov recursion relations in \cite{Kos:2013tga}. Thus, for a given $\Delta$ and $\ell$, we can expand the blocks and the integration measures at small $r$ to some order $p_\text{max}$, perform the $r$ integrals exactly, and then perform the remaining $\eta$ integral numerically. The error from this expansion scales like $(2-\sqrt{3})^{p_\text{max}}$, where $2-\sqrt{3}\approx .268$ is the maximum value of $r$ in $D_1$, and so this method converges quite quickly.

To implement the numerical bootstrap, we use the dual formulation of \eqref{eq:primal_problem}. Each constraint in the primal problem becomes a variable in the dual problem; we denote the variables corresponding to the crossing constraints by $\alpha^{(r)}_{p,q}$, and the variable corresponding to the integral constraint by $\alpha^0$. Each variable in the primal problem becomes a constraint in the dual problem; the constraint corresponding to $\lambda^2_\mathcal{M}$ for $\mathcal{M}\neq \mathcal{M}^*$ is
\begin{equation}
    \alpha^0 I[\mathfrak{G}_\mathcal{M}] + \sum_{r,p,q} \alpha^{(r)}_{p,q} V^{(r)}_{\mathcal{M};p,q} \ge 0\,,
\end{equation}
where the sum is over $r =1,2$ and the values of $p$ and $q$ given by \eqref{eq:independent_crossing}. The constraint corresponding to $\lambda^2_{\mathcal{M}^*}$ is
\begin{equation}
    \alpha^0 I[\mathfrak{G}_{\mathcal{M}^*}]+\sum_{r,p,q}\alpha^{(r)}_{p,q} {V}^{(r)}_{\mathcal{M}^*;p,q} = s \quad \text{where}\quad s =\begin{cases} 1 & \text{for maximization},\\ -1 & \text{for minimization}.\end{cases}
\end{equation}
The parts of the constraints in \eqref{eq:primal_problem} that do not depend on the unknown OPE coefficients (i.e., the terms proportional to $\lambda^2_\text{Stress}$, $\lambda^2_{(B,2)}$, and $\lambda^2_{(B,+)}$, as well as the right-hand side of the integral constraint) enter into the objective function in the dual problem. In full, the problem dual to \eqref{eq:primal_problem} is
\begin{equation}\label{eq:dual_problem}
\begin{aligned}
    & \text{max/min} && -s\left(\alpha^0 I[\mathfrak{G}_\text{Id}] + \sum_{\substack{\mathcal{M}=(B,2),\\(B,+),\text{Stress}}}\left\lbrack \alpha^0 I[\mathfrak{G}_\mathcal{M}] + \sum_{r,p,q} \alpha^{(r)}_{p,q} V^{(r)}_{\mathcal{M};p,q}\right\rbrack + \alpha^0 \frac{64}{\pi^2 c_T^2} \frac{\partial^4 F}{\partial m_+^2 \partial m_-^2}\right) \\
    & \text{s.t.} &&
    \begin{aligned}[t]
        & \alpha^0 I[\mathfrak{G}_\mathcal{M}] + \sum_{r,p,q} \alpha^{(r)}_{p,q} V^{(r)}_{\mathcal{M};p,q} \ge 0 \qquad \forall \mathcal{M}\neq\mathcal{M}^*, \\
        & \alpha^0 I[\mathfrak{G}_{\mathcal{M}^*}] + \sum_{r,p,q} \alpha^{(r)}_{p,q} V^{(r)}_{\mathcal{M}^*;p,q} = s\,.
    \end{aligned}
\end{aligned}
\end{equation}

This is a linear program that can be solved using \texttt{SDPB} \cite{Simmons-Duffin:2015qma,Landry:2019qug}, provided we first truncate the infinite set of multiplets $\mathcal{M}\neq \mathcal{M}^*$ to a finite set. We impose two truncations: first, we only include semishort and long multiplets up to some maximum spin $\ell_\text{max}$. This is standard practice for the numerical bootstrap, and one can check after the fact that the extremal functional also satisfies the constraints corresponding to higher-spin multiplets. In addition, for the long multiplets of some spin $\ell$, we only include scaling dimensions in the range $\Delta\in [\ell + 1, \ell + \tau_\text{max}]$ for some maximum twist $\tau_\text{max}$,\footnote{As discussed in the 4d case \cite{Chester:2021aun,Chester:2023ehi}, the integration region $D_1$ leads to block integrals proportional to $(-1)^{\ell/2}$ at large $\Delta$. Since the block integrals grow more quickly than $V^{(r)}_{(A,0)_{\Delta,\ell};p,q}$ at large $\Delta$, positivity at large $\Delta$ would enforce $\alpha^0 = 0$ if we set $\tau_\text{max}$ arbitrarily large. In Appendix~\ref{app:boot_numerics}, we summarize the resolution to this problem explained in \cite{Chester:2021aun,Chester:2023ehi}. In practice, as long as we judiciously choose $\tau_\text{max}$, we obtain correct bounds that make nontrivial use of the integral constraint.} and we discretize this range with some spacing $\Delta_\text{sp}$. The discretization has a minimal effect, as we can see by varying $\Delta_\text{sp}$; this is shown in detail in Appendix~\ref{app:boot_numerics}.

\subsection{Results}\label{sec:results}

\begin{figure}[t]
	\centering
	\begin{subfigure}[t]{0.48\linewidth}%
		\centering
		\includegraphics[width=\linewidth]{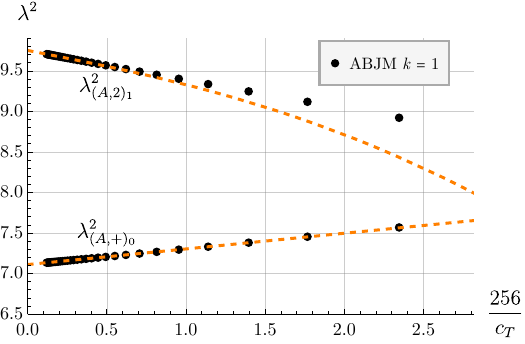}
		\caption{The $k = 1$ ABJM theory.}
	\end{subfigure}%
	\hspace{.02\linewidth}%
	\begin{subfigure}[t]{0.48\linewidth}%
		\centering
		\includegraphics[width=\linewidth]{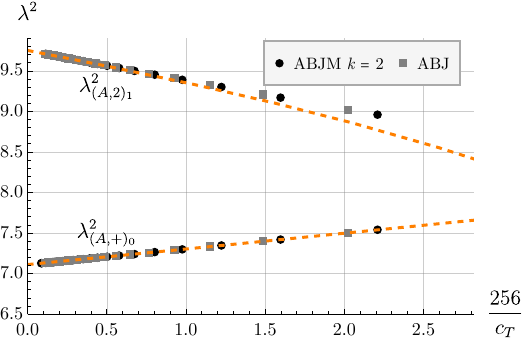}
		\caption{The $k = 2$ ABJM and ABJ theories.}
	\end{subfigure}
	\caption{The bootstrap results (points) and the large-$c_T$ expansion up to $\mathcal{O}(c_T^{-7/3})$ (dashed lines) for the squared OPE coefficients $\lambda^2_{(A,+)_{0}}$ and $\lambda^2_{(A,2)_{1}}$. The points are much larger than the actual width of the bootstrap bounds.}
	\label{fig:all_ope}
\end{figure}

Using the bootstrap method discussed in Section~\ref{sec:bootstrap_setup}, we can derive upper and lower bounds for the semishort OPE coefficients for any value of $N$ in the $k = 1,2$ ABJM theories and in the ABJ theory. We will focus on the spin-0 and spin-1 coefficients $\lambda^2_{(A,+)_{0}}$ and $\lambda^2_{(A,2)_{1}}$. In Figure~\ref{fig:all_ope}, we plot our bootstrap bounds and the large-$c_T$ expansion \eqref{strong}. At large $c_T$ we see that our bounds agree with the expansions.

\begin{figure}[t]
	\centering
	\begin{subfigure}[t]{0.48\linewidth}%
		\centering
		\includegraphics[width=\linewidth]{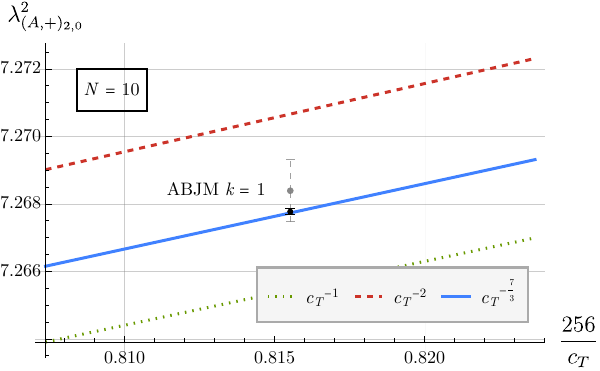}
		\caption{}
	\end{subfigure}%
	\hspace{.02\linewidth}%
	\begin{subfigure}[t]{0.48\linewidth}%
		\centering
		\includegraphics[width=\linewidth]{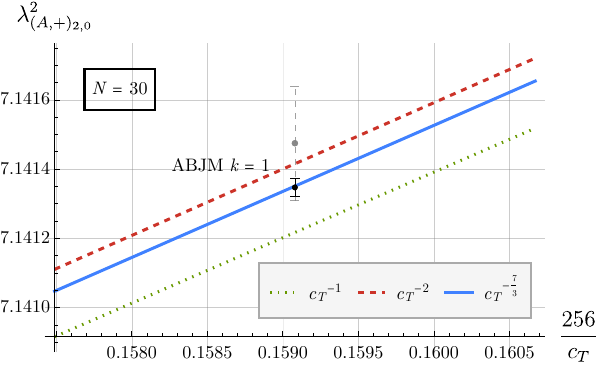}
		\caption{}
	\end{subfigure}\\[1em]
	\begin{subfigure}[t]{0.48\linewidth}%
		\centering
		\includegraphics[width=\linewidth]{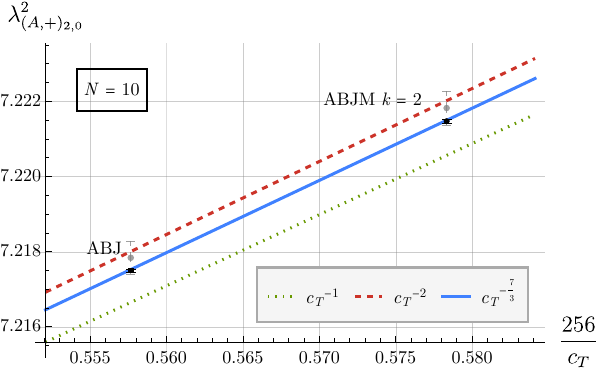}
		\caption{}
	\end{subfigure}%
	\hspace{.02\linewidth}%
	\begin{subfigure}[t]{0.48\linewidth}%
		\centering
		\includegraphics[width=\linewidth]{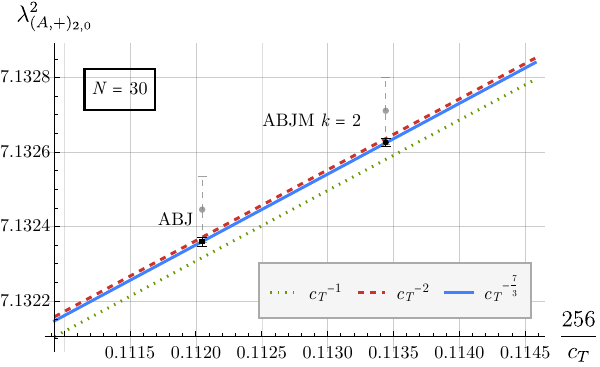}
		\caption{}
	\end{subfigure}
	\caption{Our bootstrap bounds on $\lambda^2_{(A,+)_{0}}$ for $k=1,2$ and $N=10,30$, both with the integral constraint (black) and without (light gray), compared with \eqref{strong}. The tighter bounds obtained using the integral constraint agree with the large-$c_T$ expansion once we include the $c_T^{-7/3}$ term coming from the $D^6 R^4$ correction in \eqref{cPlanck}.}
	\label{fig:d6r4_spin0}
\end{figure}

As we can see in Figure~\ref{fig:island_n2_compare}, the bounds we obtain using the mixed-mass integral constraint \eqref{con3d2} are significantly more precise than those obtained in previous studies not using this constraint (e.g. \cite{Agmon:2017xes,Agmon:2019imm}). In Figure~\ref{fig:d6r4_spin0}, we show this in another way by plotting our bounds on $\lambda^2_{(A,+)_{0}}$ at $N = 10$ and $30$ and $k = 1,2$ and comparing them with \eqref{strong}, including successive terms up to $\mathcal{O}\left(c_T^{-7/3}\right)$. We see that the integral constraint makes our bounds narrow enough to be sensitive to the $D^6 R^4$ correction that enters at $\mathcal{O}(c_T^{-7/3})$.

Since the bounds on $\lambda^2_{(A,+)_0}$ are so tight once we include the integral constraint, we can even use them to estimate the power and coefficient of the N$^3$LO correction to this OPE coefficient. We know that this correction comes from the $D^6 R^4$ term, so this will serve as a test of what we could learn in this case from the bootstrap. 

\begin{figure}[t]
	\centering
	\begin{subfigure}[t]{0.48\linewidth}%
		\centering
		\includegraphics[width=\linewidth]{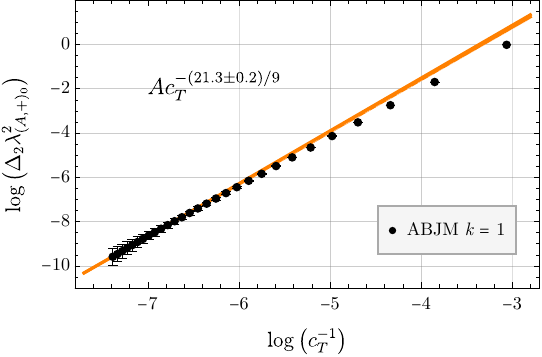}
		\caption{The N$^3$LO correction to $\lambda^2_{(A,+)_0}$ is estimated to scale like $c_T^{-7/3}$, as we expect for the term coming from a $D^6R^4$ contact diagram.}
        \label{fig:d6r4_slope}
	\end{subfigure}%
	\hspace{.02\linewidth}%
	\begin{subfigure}[t]{0.48\linewidth}%
		\centering
		\includegraphics[width=\linewidth]{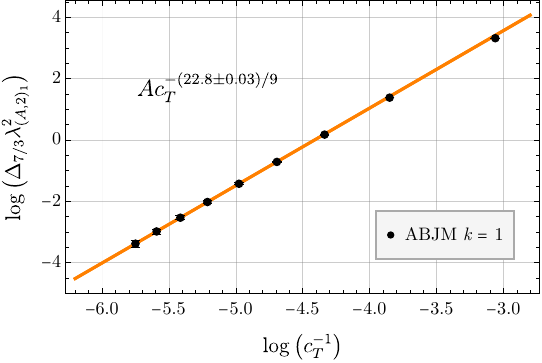}
		\caption{The N$^4$LO correction to $\lambda^2_{(A,2)_1}$ is estimated to scale like $c_T^{-23/9}$, as we expect for the term coming from a $D^8R^4$ contact diagram.}
        \label{fig:d8r4_slope}
	\end{subfigure}
	\caption{We estimate the exponents of the N$^3$LO and N$^4$LO terms in the large-$c_T$ expansions of the semishort OPE coefficients of ABJ(M) theory. In each case, we find the expected value after rounding to the nearest multiple of $1/9$.}
	\label{fig:slopes}
\end{figure}

Let us define the residual $\Delta_r \lambda^2$ as the absolute difference between an OPE coefficient $\lambda^2$ and its large-$c_T$ expansion up to order $c_T^{-r}$. In Figure~\ref{fig:d6r4_slope}, we plot $\Delta_2 \lambda^2_{(A,+)_0}$ for the $k = 1$ ABJM theory as a function of $c_T^{-1}$. By plotting on logarithmic axes and fitting for the slope, and then rounding to the nearest multiple of $1/9$, we indeed find that the N$^3$LO term in this OPE coefficient should scale like $c_T^{-7/3}$.

\begin{figure}
	\centering
	\begin{subfigure}[t]{0.48\linewidth}%
		\centering
		\includegraphics[width=\linewidth]{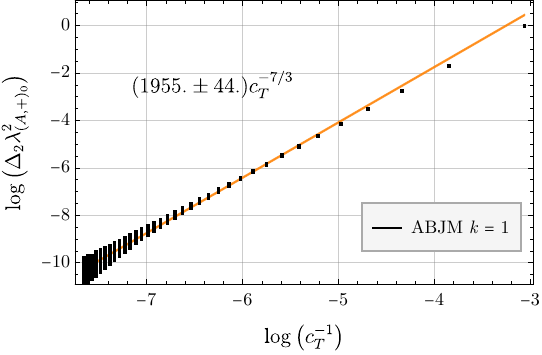}
		\caption{The estimate for the $k = 1$ ABJM theory.}
	\end{subfigure}%
	\hspace{.02\linewidth}%
	\begin{subfigure}[t]{0.48\linewidth}%
		\centering
		\includegraphics[width=\linewidth]{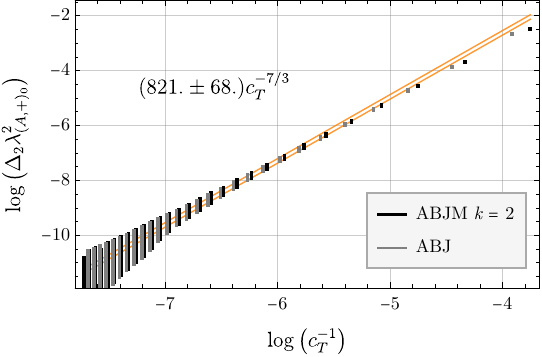}
		\caption{The estimate for the $k = 2$ ABJM theory and the ABJ theory.}
	\end{subfigure}
	\caption{The orange bars show the deviation between our bootstrap results and the large-$c_T$ expansion at $\mathcal{O}\left(c_T^{-2}\right)$ for $\lambda^2_{(A,+)_0}$. For both $k = 1$ and $k = 2$, we plot $c_T^{-7/3}$ corrections that fit within the bootstrap bounds for $N\ge 10$. This provides a rough estimate of the coefficient of this correction.}
	\label{fig:d6r4_coeff}
\end{figure}

Once we assume that this term scales as $c_T^{-7/3}$, we can use our bootstrap bounds to estimate its coefficient. In Figure~\ref{fig:d6r4_coeff}, we plot our bounds on $\Delta_2 \lambda^2_{(A,+)_0}$ as a function of $c_T$ on logarithmic axes, for both the $k = 1$ ABJM theory and the $k = 2$ ABJ(M) theories. We then plot lines corresponding to the minimum and maximum coefficients for which the large-$c_T$ expansion up to the $c_T^{-7/3}$ term would lie inside all the bounds with $N\ge 10$. In both cases, our estimates agree with the known values:
\begin{equation}
    \begin{aligned}
        k = 1&:\qquad \text{Estimated: }{-(1955\pm 44)c_T^{-7/3}}, & &\text{Actual: }{-1961.11c_T^{-7/3}} \\
        k = 2&:\qquad \text{Estimated: }{-(821\pm 68)c_T^{-7/3}}, && \text{Actual: }{-778.27c_T^{-7/3}}
    \end{aligned}
\end{equation}

Since this method is able to accurately extract the known $c_T^{-7/3}$ correction coming from the protected $D^6 R^4$ term in M-theory, we can equally well apply it to extract the $c_T^{-23/9}$ correction coming from the unprotected $D^8 R^4$ term. However, since the large-$c_T$ expansion up to order $c_T^{-7/3}$ lies within the bootstrap bounds for $\lambda^2_{(A,+)_0}$ for most values of $N$ (see Figure~\ref{fig:d6r4_spin0}), we cannot estimate the next correction to this OPE coefficient. In Figure~\ref{fig:d8r4_slope}, we instead use $\Delta_{7/3} \lambda^2_{(A,2)_1}$ to estimate how the next correction to $\lambda^2_{(A,2)_1}$ scales. We only use $2\leq N\leq 10$ for this estimation, since our bounds on $\lambda^2_{(A,2)_1}$ are weaker than those on $\lambda^2_{(A,+)_0}$ (see Figure~\ref{fig:island_n2_compare} for an example) and at larger values of $N$, $\Delta_{7/3} \lambda^2_{(A,2)_1}$ is poorly constrained. Nevertheless, with these first few values of $N$, we can estimate that the next correction to $\lambda^2_{(A,2)_1}$ after the $c_T^{-7/3}$ term scales like $c_T^{-23/9}$, as we expect for the contribution from a $D^8 R^4$ contact term.

\begin{figure}
	\centering
	\begin{subfigure}[t]{0.48\linewidth}%
		\centering
		\includegraphics[width=\linewidth]{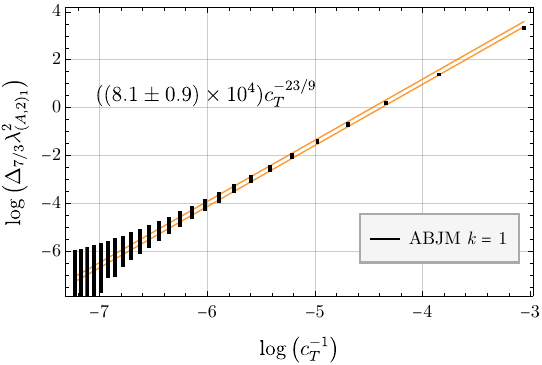}
		\caption{The estimate for the $k = 1$ ABJM theory.}
	\end{subfigure}%
	\hspace{.02\linewidth}%
	\begin{subfigure}[t]{0.48\linewidth}%
		\centering
		\includegraphics[width=\linewidth]{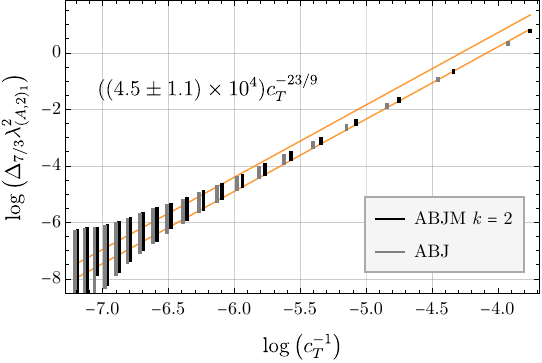}
		\caption{The estimate for the $k = 2$ ABJM theory and the ABJ theory.}
	\end{subfigure}
	\caption{The orange bars show the deviation between our bootstrap results and the large-$c_T$ expansion at $\mathcal{O}\left(c_T^{-7/3}\right)$ for $\lambda^2_{(A,2)_1}$. For both $k = 1$ and $k = 2$, we plot $c_T^{-23/9}$ corrections that fit within the bootstrap bounds for $N\ge 10$. This provides a rough estimate of the coefficient of this correction.}
	\label{fig:d8r4_coeff}
\end{figure}

We can then use our bootstrap bounds to estimate the coefficient of $c_T^{-23/9}$ in the expansion of $\lambda^2_{(A,2)_1}$, just as we estimated the coefficient of $c_T^{-7/3}$ in the expansion of $\lambda^2_{(A,+)_0}$. We show the results in Figure~\ref{fig:d8r4_coeff}. Since the $D^8 R^4$ term in M-theory is unprotected, we have no analytic prediction to compare with in this case, and the coefficients given in Figure~\ref{fig:d8r4_coeff} should be considered predictions to be compared with future work. 

As discussed in Section \ref{sec:large_ct}, the ratio between the $k = 1$ and $k = 2$ coefficients should be $2^{14/9} \approx 2.9$. The ratio of our estimates is in the range $[1.3,2.6]$, which is in the ballpark of the predicted value. The discrepancy is likely due to the arbitrary cutoff $N\ge 10$ employed in the analysis of Figure \ref{fig:d8r4_coeff}. More precise data will be needed to make better estimates in the future.

\section{Discussion}
\label{disc}

In this paper we combined the numerical conformal bootstrap with constraints from supersymmetric localization to non-perturbatively study the stress tensor correlator in 3d $\mathcal{N}=8$ ABJ(M) theory for finite $N$ and $k$. The localization constraints include both the known exact values of short OPE coefficients, which are simple to input and have been used in previous work \cite{Agmon:2017xes,Agmon:2019imm}, and an integral constraint, which is non-trivial to input and is the main technical step of this paper (see also a similar analysis for $\mathcal{N} = 4$ super-Yang-Mills in \cite{Chester:2021aun,Chester:2023ehi}). We computed precise islands for the semishort OPE coefficients squared $\lambda^2_{(A,+)_0}$ and $\lambda^2_{(A,2)_1}$. Our islands for $\lambda^2_{(A,+)_0}$ match the protected $D^6R^4$ correction from a previous analytic prediction in the large $c_T\sim N^{3/2}$ regime. Our islands for $\lambda^2_{(A,2)_1}$ were furthermore sensitive to the subsequent $1/c_T$ correction. We thus used our bounds on this coefficient to tentatively predict that the next correction scales as $c_T^{-23/9}$, implying that it comes from the $D^8R^4$ vertex, and to give an estimate for its coefficient. We found that the ratio of the coefficients between the $k=2$ and $k=1$ theories is roughly consistent with $2^{-14/9}$, which is expected for a $D^8R^4$ contact term for M-theory on AdS$_4\times S^7/\mathbb{Z}_k$.

To make further progress we would need even more precise bounds, which we can obtain in at least two different ways. One method would be to derive the integral constraint from $\partial_b^2\partial_{m_\pm}^2F\vert_{m_\pm=0,b=1}$, where $b$ parameterizes the squashing of the three-sphere. This derivative can also be computed using localization \cite{Hama:2011ea,Chester:2020jay,Chester:2021gdw}, and so it is possible that this would give us an independent constraint.\footnote{The constraint corresponding to the other nonvanishing fourth-order derivative, $\partial_b^4F\vert_{m_\pm=0,b=1}$, was shown in \cite{Bomans:2021ldw} to be redundant.} If it is independent, then this along with our numerical estimate for $\lambda^2_{(A,2)_1}$ at order $1/c_T^{23/9}$ will be sufficient to fix all the coefficients of the full holographic correlator at $D^8R^4$, and thus the M-theory S-matrix in the flat limit. 

Another approach would be to bootstrap mixed correlators with at least one of the two relevant long multiplets.\footnote{These are the double trace operator with $\Delta=2$ at $N\to\infty$, and the linear combination of the double and triple trace operators with $\Delta=3$ at $N\to\infty$.} Mixing all relevant operators was necessary to find islands in previous bootstrap studies in non-supersymmetric CFTs \cite{Kos:2014bka,Kos:2015mba}, and so it is reasonable to expect that our already-precise islands could improve substantially from imposing the additional crossing constraints of this mixed correlator system. 

Another future direction is to study $\grU(N+M)_{k}\times \grU(N)_{-k}$ ABJ(M) for general $N,k,M$ using similar methods. These theories generically have $\mathcal{N}=6$ supersymmetry. This more general parameter space includes the type IIA limit at large $k,N$ and finite $M,\lambda\equiv N/k$, and the higher-spin gravity limit at large $k,M$ and finite $N,k/M$. The stress tensor bootstrap was initiated in \cite{Binder:2020ckj}, but only two constraints were imposed via the short multiplet OPE coefficients, which was insufficient to fix the three parameters $N,k,M$. By imposing the integrated constraint $\partial_{m_+}^2\partial_{m_-}^2F(m_\pm)\vert_{m_\pm}=0$, we can hope to pin down the theory numerically. We would then be able to compare non-perturbative numerical results to the predictions of \cite{Chester:2024esn}, as well as to integrability results in the large $N$ and finite $\lambda$ regime \cite{Cavaglia:2014exa} (as was done for 4d $\mathcal{N} = 4$ SYM in \cite{Chester:2023ehi}). 

Finally, we can try to combine the numerical bootstrap with localization to non-perturbatively study 3d CFTs with even less supersymmetry. The 3d $\mathcal{N}=4$ bootstrap for correlators of the flavor multiplet was initiated in \cite{Chang:2019dzt}, but the localization value for short OPE coefficients was not imposed. If one can compute sufficiently precise islands on CFT data in this case, then this could be used to constrain the unprotected corrections to scattering of gluons on M-theory orbifolds, as discussed in \cite{Chester:2023qwo}. One could also try to bootstrap correlators of the stress tensor multiplet by combining with integral constraints from derivatives of the squashed free energy. Finally, integral constraints could also be imposed on the $\mathcal{N}=2,3$ bootstrap, which remains unexplored.\footnote{Some $\mathcal{N}=2$ bootstrap studies where integral constraints were not imposed include \cite{Baggio:2017mas,Bobev:2015vsa,Bobev:2015jxa,Chester:2015lej,Chester:2015qca}.}

\section*{Acknowledgments} 

We thank Luis Fernando Alday, David Simmons-Duffin, and Tomoki Nosaka for useful discussions. RD and SSP are supported in part by the U.S. Department of Energy under Award No. DE-SC0007968\@.  RD is also supported in
part by an NSF Graduate Research Fellowship and a Princeton University Charlotte Elizabeth Procter Fellowship. SMC is supported by the UK Engineering and Physical Sciences Research council grant number EP/Z000106/1, and the Royal Society under the grant URF\textbackslash R1\textbackslash 221310.

\appendix

\section{Numerical bootstrap details}
\label{bootApp}

For the numerical bootstrap, we follow the algorithm outlined in Section~\ref{sec:bootstrap_setup}. In Appendix~\ref{app:boot_numerics}, we discuss details of the implementation that facilitate the inclusion of the integral constraint \eqref{con3d2}. In Appendix~\ref{app:boot_delta}, we use the extremal functionals for the upper and lower bounds on $\lambda^2_{(A,+)_{0}}$ to study the scaling dimensions of the primaries of the lowest spin-0 long supermultiplets.

\subsection{Implementation}\label{app:boot_numerics}

As discussed in Section~\ref{sec:bootstrap_setup}, we use the linear programming approach to the numerical bootstrap. For every putative operator in the spectrum, we have a positivity constraint on the functional $\alpha$. In order to obtain completely rigorous bounds, we would need to impose positivity for the semishort multiplets at all spins $\ell \geq 0$, and for the long multiplets at all even spins $\ell$ and all $\Delta \ge \ell + 1$ (the unitarity bound). In practice, we truncate the spins at some $\ell_\text{max}$, we use a grid of scaling dimensions for the long operators with some spacing $\Delta_\text{sp}$, and we truncate the long operator scaling dimensions at $\Delta - \ell \leq \tau_\text{max}$.

The truncation of the spins is the least problematic; this is done in all numerical bootstrap studies, and in practice once $\ell_\text{max}$ is taken to be high enough, the resulting functionals are in fact positive for all spins, so that raising $\ell_\text{max}$ further has no effect.

\begin{figure}[t]
	\centering
	\begin{subfigure}[t]{0.48\linewidth}%
		\centering
		\includegraphics[width=\linewidth]{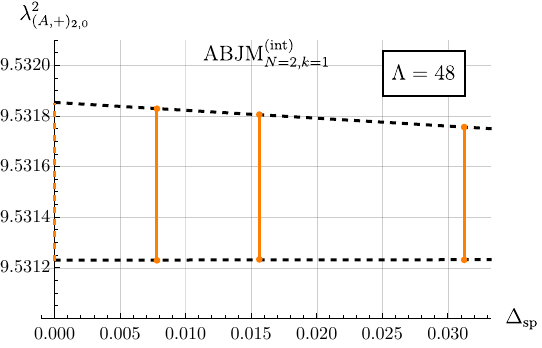}
		\caption{The $N = 2$, $k = 1$ ABJM theory.}
	\end{subfigure}%
	\hspace{.02\linewidth}%
	\begin{subfigure}[t]{0.48\linewidth}%
		\centering
		\includegraphics[width=\linewidth]{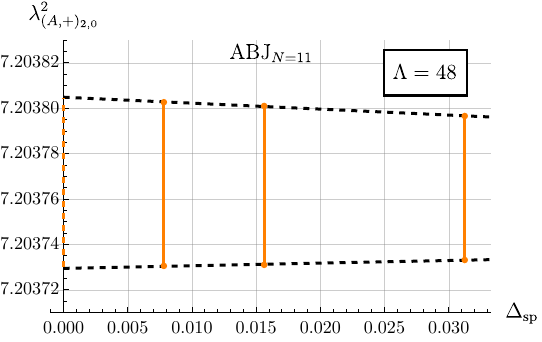}
		\caption{The $N = 11$ ABJ theory.}
	\end{subfigure}
	\caption{The numerical bootstrap bounds on $\lambda^2_{(A,+)_{2,0}}$ for two example theories as a function of the grid spacing $\Delta_\text{sp}$ for long scaling dimensions. There is only slight dependence on $\Delta_\text{sp}$, and we can reliably extrapolate $\Delta_\text{sp} \to 0$.}
	\label{fig:spacing}
\end{figure}

The truncation of the continuous set of long operator scaling dimensions to a grid with spacing $\Delta_\text{sp}$ is well-motivated: clearly as we take $\Delta_\text{sp}\to 0$ we should recover the continuous limit. For any finite $\Delta_\text{sp}$, the bounds we obtain will be slightly tighter than in the $\Delta_\text{sp} \to 0$ limit. In Figure~\ref{fig:spacing}, we check that the discrepancy is indeed slight. For the plots in Section~\ref{sec:results}, we fit a linear function of $\Delta_\text{sp}$ to bounds obtained at $\Delta_\text{sp}\in \left\lbrace \frac{1}{32},\frac{1}{64},\frac{1}{128}\right\rbrace$ and then extrapolate $\Delta_\text{sp}\to 0$. This extrapolation has no significant effect on any of our conclusions, and only serves to make the bounds more conceptually sound.

The most subtle aspect of our truncation procedure is the imposition of a cutoff $\tau_\text{max}$ on the long multiplet twists. This procedure is important for the integral constraint to have an effect; if $\tau_\text{max}$ is too large, then the integral constraint ceases to improve the bounds relative to those of \cite{Agmon:2019imm} obtained without this constraint. This is related to a difference in scaling between the conformal block derivatives and integrals of the blocks; see \cite{Chester:2021aun,Chester:2023ehi}.

In \cite{Chester:2023ehi}, it is shown for the 4d $\mathcal{N} = 4$ superconformal bootstrap that $\tau_\text{max}$ can be taken increasingly large as we increase $\Lambda$, so that as $\Lambda\to\infty$ we obtain rigorously positive functionals. We have checked that this holds in the 3d $\mathcal{N} = 8$ case as well. Alternatively, it is shown in \cite{Chester:2021aun} that if we impose the integral constraint twice using two different domains of integration, we can find rigorously positive functionals. We have also checked that this same method works in our case, but numerically it is more intensive and less stable. Thus, in this paper we work with $\tau_\text{max} = 64$, and appeal to the arguments of \cite{Chester:2023ehi} to reason that our bounds are valid.

Once we make all these truncations and construct a linear program, we use SDPB \cite{Simmons-Duffin:2015qma} to solve it with extended precision. In Table~\ref{tab:numerics}, we summarize the parameters of our numerical setup and the SDPB solver, and the values we use in this paper.

\begingroup
\renewcommand{\arraystretch}{1.1}
\begin{table}
	\centering
	\begin{tabular}{lc}
		\toprule
		Parameter & Value \\
		\midrule
		$\Lambda$ & 48 \\
		$r_\text{max}$ & 100 \\
		$\ell_\text{max}$ & 52 \\
		$\Delta_\text{sp}$ & $\left\lbrace \frac{1}{32},\frac{1}{64},\frac{1}{128}\right\rbrace$ \\
		$h_\text{max}$ & 64 \\
		\midrule
		\texttt{precision} & 512 \\
		\texttt{dualityGapThreshold} & $10^{-30}$ \\
		\texttt{primalErrorThreshold} & $10^{-30}$ \\
		\texttt{dualErrorThreshold} & $10^{-30}$ \\
		\texttt{feasibleCenteringParameter} & $0.1$ \\		
		\texttt{infeasibleCenteringParameter} & $0.3$ \\
		\texttt{stepLengthReduction} & $0.7$ \\
		\texttt{maxComplementarity} & $10^{100}$ \\
		\bottomrule
	\end{tabular}
	\caption{The parameters we use in constructing linear programs (above) and in solving them with SDPB \cite{Simmons-Duffin:2015qma} (below).}
	\label{tab:numerics}
\end{table}
\endgroup

\subsection{Extremal functionals}\label{app:boot_delta}

After solving the linear programs discussed in the previous section to optimality, the optimal value provides bounds on the targeted OPE coefficients while the optimal solution provides an extremal functional $\alpha$ that can be used to estimate other CFT data \cite{El-Showk:2012vjm}. Here we will study the extremal functional associated to the lower bound on $\lambda^2_{(A,+)_{2,0}}$, which has been found in past studies to correspond most closely to ABJM theory.

\begin{figure}[t]
	\centering
	\begin{subfigure}[t]{0.48\linewidth}%
		\centering
		\includegraphics[width=\linewidth]{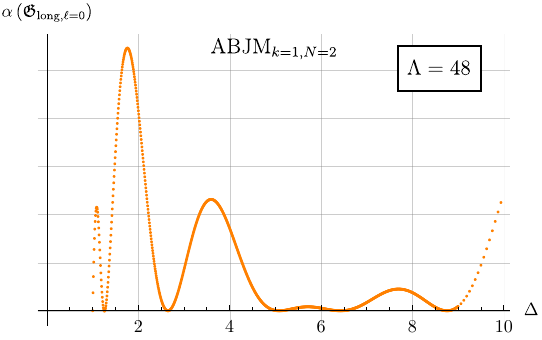}
		\caption{The $N = 2$, $k = 1$ ABJM theory.}
	\end{subfigure}%
	\hspace{.02\linewidth}%
	\begin{subfigure}[t]{0.48\linewidth}%
		\centering
		\includegraphics[width=\linewidth]{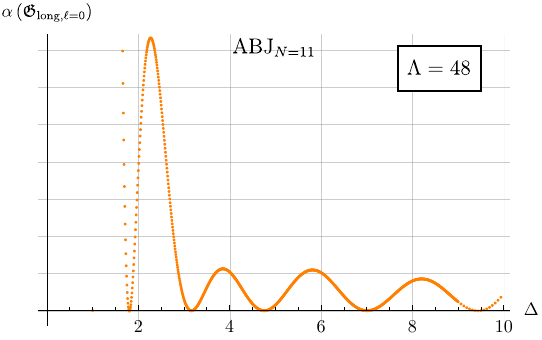}
		\caption{The $N = 11$ ABJ theory.}
	\end{subfigure}
	\caption{Extremal functionals  for the minimum value of $\lambda^2_{(A,+)_{2,0}}$ applied to the spin-0 long superblock as a function of $\Delta$. The zeroes of these functions are estimates for the scaling dimensions of the primary operators of spin-0 long supermultiplets in the spectrum.}
	\label{fig:functionals}
\end{figure}

By applying the extremal functional to the long superblocks, we can estimate the scaling dimensions of the primary operators in long supermultiplets. In Figure~\ref{fig:functionals}, we give representative examples of extremal functionals acting upon the spin-0 long superblock. The zeroes of these functionals are the estimated scaling dimensions of the primary operators of spin-0 long supermultiplets.

\begin{figure}[t]
	\centering
	\begin{subfigure}[t]{0.48\linewidth}%
		\centering
		\includegraphics[width=\linewidth]{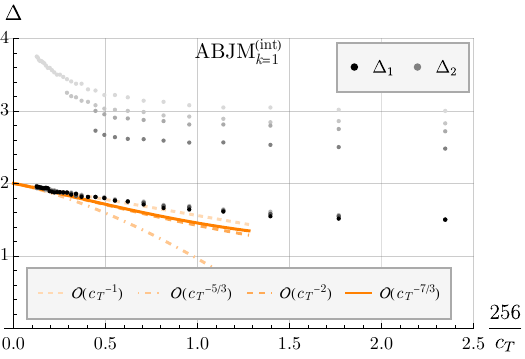}
		\caption{The $k = 1$ ABJM theory.}
	\end{subfigure}%
	\hspace{.02\linewidth}%
	\begin{subfigure}[t]{0.48\linewidth}%
		\centering
		\includegraphics[width=\linewidth]{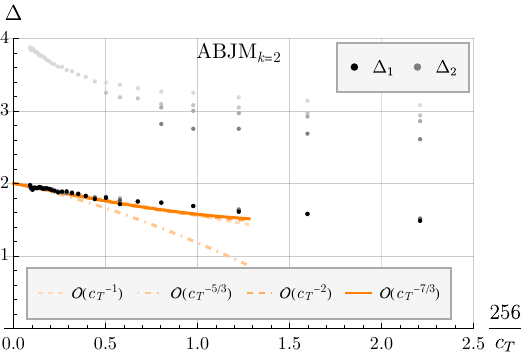}
		\caption{The $k = 2$ ABJM theory.}
	\end{subfigure}\\
	\begin{subfigure}[t]{0.48\linewidth}%
		\centering
		\includegraphics[width=\linewidth]{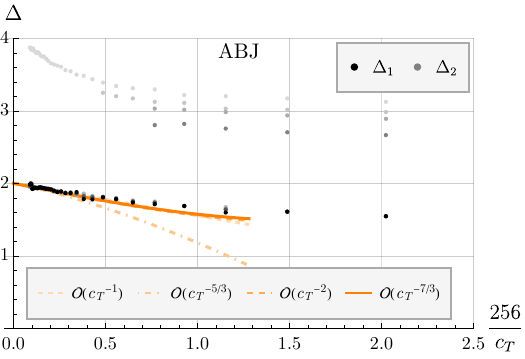}
		\caption{The ABJ theory.}
	\end{subfigure}
	\caption{We use the zeroes of extremal functionals like those in Figure~\ref{fig:functionals} to estimate the scaling dimensions of the primary operators in the lowest two spin-0 long supermultiplets. We plot the estimates at $\Lambda = 24$, 36, 48, and the $\Lambda\to\infty$ extrapolation in increasingly dark points. For the lowest of these, which has twist 2 in the large-$c_T$ limit, we compare with a large-$c_T$ expansion at successive orders and find good agreement.}
	\label{fig:efm_dimensions}
\end{figure}

In Figure~\ref{fig:efm_dimensions}, we use these zeroes to estimate the scaling dimensions of the two lowest spin-0 long supermultiplets. The lower of these operators has twist 2 in the $N\to\infty$ limit, and the large-$c_T$ expansion of its scaling dimension begins \cite{Chester:2018aca,Alday:2022rly,Alday:2021ymb}
\begin{equation}
	\Delta_1 = 2 -\frac{1120}{\pi ^2 c_T } -\frac{71680\times 6^{1/3}}{\pi ^{8/3} c_T^{5/3} k^{2/3}} + \frac{a_k}{c_T^2} + \frac{716800 (\frac{2}{3})^{1/3}}{\pi ^{10/3} c_T^{7/3} k^{4/3}} + \cdots \,,
\end{equation}
where
\begin{equation}
a_1 = -\frac{1120 \left(854313 \pi ^2-11803792\right)}{1287 \pi ^4}\,, \qquad a_2 = -\frac{8960 \left(39408 \pi ^2-737737\right)}{1287 \pi ^4}\,.
\end{equation}
We compare this expansion with the data for all three theories we consider, and find good agreement in each case.

The next-lowest operator is already much more difficult. In Figure~\ref{fig:efm_dimensions} we plot its dimension estimated with derivative orders $\Lambda = 24$, 36, and 48, showing that it is still far from being converged in $\Lambda$. In addition, at larger values of $N$ and $\Lambda$ we find that the extremal functionals applied to the spin-0 long superblock only have one minimum close to 0, so we cannot estimate the dimension of the second operator in these cases.

Thus, although at $\Lambda = 24$ we see clearly in Figure~\ref{fig:efm_dimensions} that the scaling dimension is approaching 4, it is plausible that as $\Lambda\to\infty$ it is actually approaching 3. More precision will be needed to discern between these possibilities, and we leave this question for future work.

\section{Localization}\label{app:localization}

In \cite{Nosaka:2024gle}, Nosaka impressively calculates the exact partition function of the mass-deformed $\grU(N)_k \times \grU(N+M)_k$ ABJ(M) model. The result is given in terms of some rather involved recursion relations. Here we summarize the recursion relations in the cases $(k,M) \in \{(1,0), (2,0),(2,1)\}$ that we use in this paper.

Let $Z_{k,M,N}$ be the partition function of the mass-deformed $\grU(N)_k \times \grU(N+M)_{-k}$ model. When $k = 1$, the recursion relation from \cite{Nosaka:2024gle} is
\begin{equation}\label{eq:recursion_k1}
\begin{split}
	Z_{1,0,N} = \frac{1}{I_{1,0,N} + I_{1,0,-N}}\Bigg\lbrack &\sum_{n=0}^{N-1} \left(i \exp\left(\frac{i\pi}{2}m_+ m_-\right)\right)^{2n+1-N} Z_{1,0,n} Z_{1,0,N-1-n} \\
	&- \sum_{n=1}^{N - 1} I_{1,0,2n-N} Z_{1,0,n} Z_{1,0,N-n}\Bigg\rbrack\,,
\end{split}
\end{equation}
where
\begin{equation}
	I_{k,\ell,N} = \exp\left(\frac{N\pi}{2}(m_+ - m_-)\right) - (-1)^N \exp\left(\frac{2\pi i \ell}{k}\right)\exp\left(\frac{N\pi}{2}(m_+ + m_-)\right)\,.
\end{equation}
This allows us to compute $Z_{1,0,N}$ for any $N\ge 2$ from the base cases
\begin{equation}
\begin{split}
	Z_{k,M,1} &= \frac{1}{4k\cosh(\pi m_+/2)\cosh(\pi m_-/2)}\,,\\
	Z_{k,M,0} &= i^{M(M-2)/2}\exp\left(-i\pi\frac{M(M^2-1)}{6k}\right)k^{-M/2}\prod_{s=1}^M \prod_{r=s+1}^M 2\sin\left(\frac{\pi(r-s)}{k}\right)\,.
\end{split}
\end{equation}

\begingroup
\renewcommand{\arraystretch}{1.5}
\begin{table}[t]
	\centering
	\begin{tabular}{cccc}
	\toprule
	$N$ & $\text{ABJM}^\text{(int)}_{k=1}$ & $\text{ABJM}_{k=2}$ & ABJ \\
	\midrule
	$2$ & $\frac{64}{3}$ & $\frac{128}{3}$ & $\frac{32 \left(9 \pi ^2-80\right)}{3 \left(\pi ^2-8\right)}$\\
$3$ & $\frac{16 (10 \pi -31)}{\pi -3}$ & $\frac{7136-720 \pi ^2}{30-3 \pi ^2}$ & $\frac{32 \left(49 \pi ^2-480\right)}{5 \pi ^2-48}$\\
$4$ & $\frac{7136-720 \pi ^2}{30-3 \pi ^2}$ & $\frac{64 \left(672-512 \pi ^2+45 \pi ^4\right)}{72-96 \pi ^2+9 \pi ^4}$ & $\frac{64 \left(14592-21464 \pi ^2+2025 \pi ^4\right)}{3 \left(480-848 \pi ^2+81 \pi ^4\right)}$\\
$5$ & $\frac{64 \left(-775-545 \pi +252 \pi ^2\right)}{-78-60 \pi +27 \pi ^2}$ & $\frac{16 \left(165912-156712 \pi ^2+14175 \pi ^4\right)}{3 \left(1032-904 \pi ^2+81 \pi ^4\right)}$ & $\frac{640 \left(31920-20200 \pi ^2+1719 \pi ^4\right)}{3 \left(6240-800 \pi ^2+17 \pi ^4\right)}$\\
	\bottomrule
	\end{tabular}
	\caption{Exact values of $c_T$ for the theories considered in this paper with $2\leq N\leq 5$.}
	\label{tab:cT}
\end{table}

\begin{table}[t]
	\centering
	\begin{tabular}{cc}
	\toprule
	$N$ & $\left(\lambda^2_{(B,2)}\right)^{\text{(int)}}$ \\
	\midrule
	$2$ & $0$\\
$3$ & $\frac{16 \left(3888-2557 \pi +420 \pi ^2\right)}{3 (31-10 \pi )^2}$\\
$4$ & $\frac{64 \left(156448-31396 \pi ^2+1575 \pi ^4\right)}{3 \left(446-45 \pi ^2\right)^2}$\\
$5$ & $\frac{4 \left(8194276+11967910 \pi -1143732 \pi ^2-3966435 \pi ^3+908334 \pi ^4\right)}{3 \left(775+545 \pi -252 \pi ^2\right)^2}$\\
	\bottomrule
	\end{tabular}
	\caption{Exact values of $\lambda^2_{(B,2)}$ for the $\text{ABJM}^\text{(int)}_{k=1}$ theory with $2\leq N\leq 5$.}
	\label{tab:lambdaB2_1}
\end{table}

\begin{table}[t]
	\centering
	\begin{tabular}{cc}
	\toprule
	$N$ & $\lambda^2_{(B,2)}$ \\
	\midrule
	$2$ & $\frac{16}{3}$\\
$3$ & $\frac{64 \left(156448-31396 \pi ^2+1575 \pi ^4\right)}{3 \left(446-45 \pi ^2\right)^2}$\\
$4$ & $\frac{8 \left(377856-1069056 \pi ^2+559480 \pi ^4-82080 \pi ^6+3645 \pi ^8\right)}{\left(672-512 \pi ^2+45 \pi ^4\right)^2}$\\
$5$ & $\frac{128 \left(3581581824-6553080000 \pi ^2+3566316232 \pi ^4-535767696 \pi ^6+24111675 \pi ^8\right)}{\left(165912-156712 \pi ^2+14175 \pi ^4\right)^2}$\\
	\bottomrule
	\end{tabular}
	\caption{Exact values of $\lambda^2_{(B,2)}$ for the $\text{ABJM}_{k=2}$ theory with $2\leq N\leq 5$.}
	\label{tab:lambdaB2_20}
\end{table}

\begin{table}[t]
	\centering
	\begin{tabular}{cc}
	\toprule
	$N$ & $\lambda^2_{(B,2)}$ \\
	\midrule
	$2$ & $\frac{16 \left(16384-3872 \pi ^2+225 \pi ^4\right)}{3 \left(80-9 \pi ^2\right)^2}$\\
$3$ & $\frac{16 \left(574464-119168 \pi ^2+6177 \pi ^4\right)}{3 \left(480-49 \pi ^2\right)^2}$\\
$4$ & $\frac{8 \left(379994112-1202230272 \pi ^2+1014713888 \pi ^4-170189424 \pi ^6+8037225 \pi ^8\right)}{\left(14592-21464 \pi ^2+2025 \pi ^4\right)^2}$\\
$5$ & $-\frac{2 \left(-123358003200+89133772800 \pi ^2+15236652320 \pi ^4-5319414560 \pi ^6+302837409 \pi ^8\right)}{15 \left(31920-20200 \pi ^2+1719 \pi ^4\right)^2}$\\
	\bottomrule
	\end{tabular}
	\caption{Exact values of $\lambda^2_{(B,2)}$ for the ABJ theory with $2\leq N\leq 5$.}
	\label{tab:lambdaB2_21}
\end{table}

\begin{table}[t]
	\centering
	\begin{tabular}{cc}
	\toprule
	$N$ & $-\frac{64}{\pi^2 c_T^2}\frac{\partial^4 F}{\partial m_+^2 \partial m_-^2}$ \\
	\midrule
	$2$ & $\frac{3}{128}$\\
$3$ & $\frac{648-216 \pi +\pi ^3}{192 (31-10 \pi )^2}$\\
$4$ & $\frac{5 \left(984-228 \pi ^2+13 \pi ^4\right)}{8 \left(446-45 \pi ^2\right)^2}$\\
$5$ & $\frac{195312+228240 \pi -40008 \pi ^2-65530 \pi ^3+14274 \pi ^4+405 \pi ^5}{256 \left(775+545 \pi -252 \pi ^2\right)^2}$\\
	\bottomrule
	\end{tabular}
	\caption{Exact values of $-\frac{64}{\pi^2 c_T^2}\frac{\partial^4 F}{\partial m_+^2 \partial m_-^2}$ for the $\text{ABJM}^\text{(int)}_{k=1}$ theory with $2\leq N\leq 5$.}
	\label{tab:rhs_1}
\end{table}

\begin{table}[t]
	\centering
	\begin{tabular}{cc}
	\toprule
	$N$ & $-\frac{64}{\pi^2 c_T^2}\frac{\partial^4 F}{\partial m_+^2 \partial m_-^2}$ \\
	\midrule
	$2$ & $\frac{3}{256}$\\
$3$ & $\frac{5 \left(984-228 \pi ^2+13 \pi ^4\right)}{8 \left(446-45 \pi ^2\right)^2}$\\
$4$ & $\frac{3 \left(2448+192 \pi ^2+306 \pi ^4-75 \pi ^6+4 \pi ^8\right)}{8 \left(672-512 \pi ^2+45 \pi ^4\right)^2}$\\
$5$ & $\frac{3 \left(20384064-49730592 \pi ^2+34282144 \pi ^4-5500008 \pi ^6+254907 \pi ^8\right)}{2 \left(165912-156712 \pi ^2+14175 \pi ^4\right)^2}$\\
	\bottomrule
	\end{tabular}
	\caption{Exact values of $-\frac{64}{\pi^2 c_T^2}\frac{\partial^4 F}{\partial m_+^2 \partial m_-^2}$ for the $\text{ABJM}_{k=2}$ theory with $2\leq N\leq 5$.}
	\label{tab:rhs_20}
\end{table}

\begin{table}[t]
	\centering
	\begin{tabular}{cc}
	\toprule
	$N$ & $-\frac{64}{\pi^2 c_T^2}\frac{\partial^4 F}{\partial m_+^2 \partial m_-^2}$ \\
	\midrule
	$2$ & $\frac{384-48 \pi ^2+\pi ^4}{8 \left(80-9 \pi ^2\right)^2}$\\
$3$ & $\frac{3 \left(7680+496 \pi ^2-287 \pi ^4+16 \pi ^6\right)}{32 \left(480-49 \pi ^2\right)^2}$\\
$4$ & $\frac{3 \left(1912320-5026560 \pi ^2+4312496 \pi ^4-725496 \pi ^6+34263 \pi ^8\right)}{16 \left(14592-21464 \pi ^2+2025 \pi ^4\right)^2}$\\
$5$ & $\frac{540933120-1008092160 \pi ^2-1649457984 \pi ^4+2488732512 \pi ^6-449611429 \pi ^8+21822048 \pi ^{10}}{512 \left(31920-20200 \pi ^2+1719 \pi ^4\right)^2}$\\
	\bottomrule
	\end{tabular}
	\caption{Exact values of $-\frac{64}{\pi^2 c_T^2}\frac{\partial^4 F}{\partial m_+^2 \partial m_-^2}$ for the ABJ theory with $2\leq N\leq 5$.}
	\label{tab:rhs_21}
\end{table}
\endgroup

To compute the constraints in Section~\ref{sec:constraints}, we do not need the exact function $Z_{1,0,N}$, only its series expansion up to terms of degree 4. When $N$ is odd, the leading term in the prefactor $(I_{1,0,N}+I_{1,0,-N})^{-1}$ in \eqref{eq:recursion_k1} is a constant, and so we can use \eqref{eq:recursion_k1} to find the coefficients up to degree $p$ in terms of those same coefficients at lower values of $N$. However, when $N$ is even the prefactor has a leading term $\mathcal{O}\left(\frac{1}{m_+ m_-}\right)$, and so in order to find the coefficients up to degree $p$ we need the coefficients up to degree $p+2$ at lower values of $N$.

For $k = 2$, the recursion relation mixes the cases of $M = 0$ and $M = 1$. Defining
\begin{equation}
	Z'_{2,1,N} = \exp\left(-\frac{3\pi i}{4}\right)
   \left(-e^{-\frac{1}{2} \pi  (m_+ + m_-)}\right)^{N}\times Z_{2,1,N}\,,
\end{equation}
we have
\begin{equation}\label{eq:recursion_k2_0}
\begin{split}
	Z_{2,0,N} = \frac{1}{I_{2,0,N} + I_{2,0,-N}}\Bigg\lbrack &\sum_{n=0}^{N-1} \left(i \exp\left(\frac{i\pi}{2}m_+ m_-\right)\right)^{2n} Z'_{2,1,n} Z'_{2,1,N-1-n} \\
	&- \sum_{n=1}^{N - 1} I_{2,0,2n-N} Z_{2,0,n} Z_{2,0,N-n}\Bigg\rbrack
\end{split}
\end{equation}
and
\begin{equation}\label{eq:recursion_k2_1}
\begin{split}
	Z'_{2,1,N} = \frac{1}{I_{2,1,N} + I_{2,1,-N}}\frac{1}{Z'_{2,1,0}}\Bigg\lbrack &\sum_{n=0}^{N-1} \left(i \exp\left(\frac{i\pi}{2}m_+ m_-\right)\right)^{-2n} Z_{2,0,n} Z_{2,0,N-1-n} \\
	&- \sum_{n=1}^{N - 1} I_{2,1,2n-N} Z'_{2,1,n} Z'_{2,1,N-n}\Bigg\rbrack\,.
\end{split}
\end{equation}
In this case, the prefactor $(I_{2,0,N}+I_{2,0,-N})^{-1}$ is $\mathcal{O}\left(\frac{1}{m_1 m_2}\right)$ when $N$ is even, and the prefactor $(I_{2,0,N}+I_{2,0,-N})^{-1}$ is $\mathcal{O}\left(\frac{1}{m_1 m_2}\right)$ when $N$ is odd. Thus, to get the expansions of $Z_{2,0,N}$ and $Z_{2,1,N}$ up to terms of degree $p$, we need the expansions at all lower values of $N$ up to degree $p+2$, regardless of the value of $N$. This means the difficulty of the calculation grows more quickly with $N$ in the $k = 2$ case.

Using these recursion relations along with \eqref{eq:m2der}, \eqref{eq:m4der}, and \eqref{eq:crossConstraints}, we can compute $c_T = 256/\lambda^2_\text{Stress}$, $\lambda^2_{(B,+)}$, and $\lambda^2_{(B,2)}$, as well as the mixed-mass derivative needed in \eqref{con3d2}. For $k = 1$, there is an important caveat in the OPE coefficients: the $\grU(N)_k \times \grU(N)_{-k}$ model flows to the product of a free theory with $c_T = 16$ and an interacting theory. We are interested in the interacting part, and so we use the following formulas from \cite{Agmon:2019imm}:
\begin{equation}
\begin{split}
	c_T^{\text{(int)}} &= c_T - 16\,,\\
	\left(\lambda^2_{(B,2)}\right)^{\text{(int)}} &= \frac{\lambda^2_{(B,2)} c_T^2 - \frac{1024}{3}c_T}{\left(c_T^{\text{(int)}}\right)^2}\,.
\end{split}
\end{equation}
These can be used along with \eqref{eq:crossConstraints} to find $\left(\lambda^2_{(B,+)}\right)^{\text{(int)}}$.

In Tables \ref{tab:cT} through \ref{tab:rhs_21}, we give the explicit values of $c_T$, $\lambda^2_{(B,+)}$, $\lambda^2_{(B,2)}$, and the mixed-mass derivative for $2\leq N\leq 5$. As an interesting aside, it is sometimes said that one never needs particularly precise values of $\pi$ in physics, because even calculating the circumference of the universe to within the radius of a hydrogen atom requires only 40 or so digits. However, calculating the short OPE coefficients of ABJM theory actually requires significantly more precision than this, due to near-cancellations which amplify roundoff error in floating-point arithmetic. For example, in the denominator of $\lambda^2_{(B,2)}$ for the $k = 1$, $N = 5$ ABJM theory, we have
\begin{equation}
	775 + 545\pi - 252\pi^2 \approx 0.028\,.
\end{equation}
These near-cancellations become even more dramatic at larger values of $N$, so that computing the same coefficient at $N = 20$ requires about 110 digits of $\pi$ in order to obtain 10 digits of precision in the answer.

These precision issues are easily surmountable, but as explained above the difficulty of using the recursion relations of \cite{Nosaka:2024gle} increases with $N$, and increases more quickly for the $k = 2$ theories. When it becomes impractical to compute the localization input exactly, we can use the following alternative approach. The mass-deformed partition function $Z(m_\pm)\equiv -\log F(m_\pm)$ can be computed in terms of the matrix model integral \cite{Kapustin:2009kz}
 \es{ZABJM}{
 & Z({m_+,m_-} )=  \int \frac{d^{N+M} \mu d^{N} \nu}{N!(N+M)!} e^{i \pi k \left[   \sum_i \nu_i^2-\sum_a \mu_a^2 \right]} \\
  &\times \frac{\prod_{a<b} \left( 4 \sinh^2 \left[ \pi (\mu_a - \mu_b) \right] \right) 
    \prod_{i<j} \left( 4 \sinh^2 \left[ \pi (\nu_i - \nu_j) \right] \right) }{\prod_{i, a} \left(
     4 \cosh \left[\pi (\mu_a - \nu_i + m_+/2  ) \right] 
     \cosh \left[\pi (\nu_i - \mu_a + m_-/2  ) \right]  \right) } \,,
 }
 where $a=1,\dots,N+M$ and $i=1,\dots,N$. For large $N$, the mass deformed partition function was computed to all orders in $1/N$ (i.e. neglecting large $N$ instanton contributions proportional to $e^{-\sqrt{kN}}$ or $e^{-\sqrt{\frac{N}{k}}}$) using the Fermi gas method, giving \cite{Marino:2011eh,Nosaka:2015iiw}
 \es{GotZABJM}{
 & Z({m_+,m_-} ) \approx e^A C^{-\frac 13} \text{Ai}\left[C^{-\frac 13} (N-B) \right] \,,\\
  C &= \frac{2}{\pi^2 k (1 + m_+^2) (1 + m_-^2)} \,, \qquad
   B = \frac{\pi^2 C}{3} - \frac{1}{6k} \left[ \frac{1}{1 + m_+^2} + \frac{1}{1 + m_-^2} \right] + \frac{k}{24} \,, \\
  A&= \frac{{\cal A}[k(1 + i m_+)] + {\cal A}[k(1 - i m_+)] +  {\cal A}[k(1 + i m_-)] + {\cal A}[k(1 - i m_-)] }{4}  \,,
 } 
 where the constant map function ${\cal A}$ is given by
\es{constantMap}{
{\cal A}(k)&=\frac{2\zeta(3)}{\pi^2k}\left(1-\frac{k^3}{16}\right)+\frac{k^2}{\pi^2}\int_0^\infty dx\frac{x}{e^{kx}-1}\log\left(1-e^{-2x}\right)\,.
}

\begin{figure}[t]
	\centering
	\begin{subfigure}[t]{0.48\linewidth}%
		\centering
		\includegraphics[width=\linewidth]{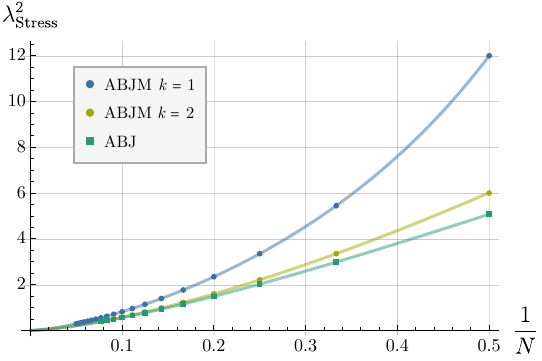}
		\caption{}
	\end{subfigure}%
	\hspace{.02\linewidth}%
	\begin{subfigure}[t]{0.48\linewidth}%
		\centering
		\includegraphics[width=\linewidth]{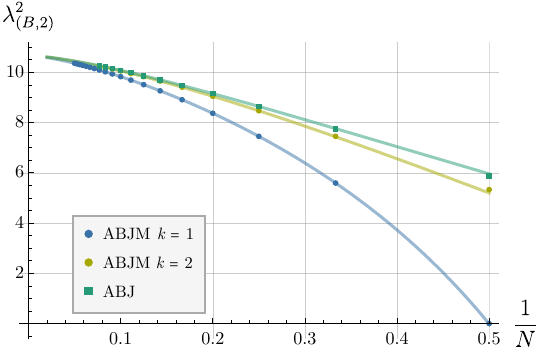}
		\caption{}
	\end{subfigure}\\[1em]
	\begin{subfigure}[t]{0.48\linewidth}%
		\centering
		\includegraphics[width=\linewidth]{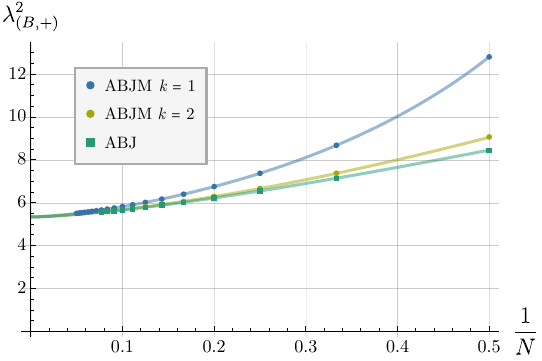}
		\caption{}
	\end{subfigure}%
	\hspace{.02\linewidth}%
	\begin{subfigure}[t]{0.48\linewidth}%
		\centering
		\includegraphics[width=\linewidth]{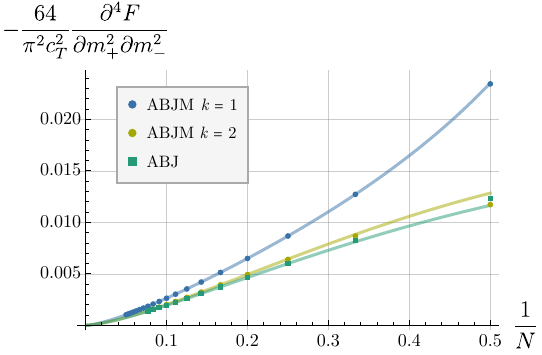}
		\caption{}
	\end{subfigure}
	\caption{The squared OPE coefficients $\lambda^2_\text{Stress}$, $\lambda^2_{(B,+)}$, $\lambda^2_{(B,2)}$, and the mixed-mass derivative of the free energy $\frac{\partial^4 F}{\partial m_+^2 \partial m_-^2}$ are computed using both the all-orders in $1/N$ expansion \eqref{GotZABJM} (dashed lines) and the exact recursion relations derived in \cite{Nosaka:2024gle}. The all-orders formula agrees with the exact result to many decimal places for $N\gtrapprox 10$.}
	\label{fig:nosaka_compare}
\end{figure}

In Figure~\ref{fig:nosaka_compare}, we plot OPE coefficients obtained using this all-orders-in-$1/N$ formula along with the exact values obtained from the recursion relations of \cite{Nosaka:2024gle}. For the $k = 1$ theory, we have computed the exact values up to $N = 20$. These values are as follows, where the bolded digits are the ones where the exact answer begins to differ from the results of the all-orders-in-$1/N$ formula:
\begin{equation}
\begin{split}
    \lambda^2_\text{Stress}(k=1,N=20) &\approx 0.290546232805027\bm{59}\,, \\
    \lambda^2_{(B,+)}(k=1,N=20) &\approx 5.5041065451068\bm{81}\,, \\
    \lambda^2_{(B,2)}(k=1,N=20) &\approx 10.3583477943142\bm{99}\,.
\end{split}
\end{equation}
In the $k = 2$ case, we have computed the exact values up to $N = 13$. For $M = 0$, the values at $N = 13$ are
\begin{equation}
\begin{split}
    \lambda^2_\text{Stress}(k=2,N=13,M=0) &\approx 0.39261716\bm{19}\,, \\
    \lambda^2_{(B,+)}(k=2,N=13,M=0) &\approx 5.5637396\bm{94}\,, \\
    \lambda^2_{(B,2)}(k=2,N=13,M=0) &\approx 10.248229\bm{82}\,;
\end{split}
\end{equation}
for $M = 1$, they are
\begin{equation}
\begin{split}
    \lambda^2_\text{Stress}(k=2,N=13,M=1) &\approx 0.38171195\bm{32}\,, \\
    \lambda^2_{(B,+)}(k=2,N=13,M=1) &\approx 5.5572900\bm{07}\,, \\
    \lambda^2_{(B,2)}(k=2,N=13,M=1) &\approx 10.259602\bm{22}\,.
\end{split}
\end{equation}
In all of these cases, the error incurred by using the all-orders-in-$1/N$ formula is much smaller than the relative precision of our bootstrap bounds. Thus, for larger values of $N$ at which this error will be even smaller, we can safely use this formula.

\bibliographystyle{ssg}
\bibliography{ABJMdraft}

\end{document}